\documentclass[%
preprint
 amsmath, amssymb, %
 aip,cha,%
groupedaddress,%
]{revtex4-1}

\usepackage{amsmath}
\usepackage{amssymb}
\usepackage{docs}%
\usepackage{bm}%
\usepackage{xcolor}
\usepackage{dcolumn}
\usepackage{graphicx}
\usepackage[colorlinks=true,linkcolor=blue]{hyperref}%
\usepackage{subcaption}
\usepackage{color}
\expandafter\ifx\csname package@font\endcsname\relax\else
 \expandafter\expandafter
 \expandafter\usepackage
 \expandafter\expandafter
 \expandafter{\csname package@font\endcsname}%
\fi
\begin{document}

\title{Deconvolutional artificial neural network models for large eddy simulation of turbulence}

\author{Zelong Yuan}%
\author{Chenyue Xie}%
\author{Jianchun Wang}%
\email[Email address for correspondence:\;]{wangjc@sustech.edu.cn}
\affiliation{\small Guangdong Provincial Key Laboratory of Turbulence Research and Applications, \\ Center for Complex Flows and Soft Matter Research, \\ Department of Mechanics and Aerospace Engineering, Southern University of Science and Technology, Shenzhen 518055, People's Republic of China}%


\date{\today}

\begin{abstract}
 Deconvolutional artificial neural network (DANN) models are developed for subgrid-scale (SGS) stress in large eddy simulation (LES) of turbulence. The filtered velocities at different spatial points are used as input features of the DANN models to reconstruct the unfiltered velocity. The grid width of the DANN models is chosen to be smaller than the filter width, in order to accurately model the effects of SGS dynamics. The DANN models can predict the SGS stress more accurately than the conventional approximate deconvolution method (ADM) and velocity gradient model (VGM) in the \emph{a prior} study: the correlation coefficients can be made larger than 99\% and the relative errors can be made less than 15\% for the DANN model. In an \emph{a posteriori} study, a comprehensive comparison of the DANN model, the implicit large eddy simulation (ILES), the dynamic Smagorinsky model (DSM), and the dynamic mixed model (DMM) shows that: the DANN model is superior to the ILES, DSM, and DMM models in the prediction of the velocity spectrum, various statistics of velocity and the instantaneous coherent structures without increasing the considerable computational cost. Besides, the trained DANN models without any fine-tuning can predict the velocity statistics well for different filter widths. These results indicate that the DANN framework with consideration of SGS spatial features is a promising approach to develop advanced SGS models in the LES of turbulence.   
\end{abstract}

\maketitle

\section{introduction} \label{sec:level1}
Large eddy simulation (LES) solves the large-scale motions of turbulence and models the effects of small scale dynamics on the large scale structures with SGS stress \cite{pope2000,sagaut2006,garnier2009,sagaut2018}. Compared to the direct numerical simulation (DNS), LES can significantly reduce the degree of freedom and predict large-scale flow structures with high accuracy\cite{lesieur1996,meneveau2000,meneveau2011,durbin2018}. Therefore, LES has been widely applied to complex turbulent flows in combustion, acoustics, astrophysics, atmospheric boundary layer mixing, \emph{et al.} \cite{pitsch2006,fureby2008,georgiadis2010,stevens2017}. Many classical closure models have been proposed to reconstruct the SGS stress, including the Smagorinsky model \cite{smagorinsky1963,lilly1967,deardorff1970}, the scale-similarity model \cite{bardina1980,liu1994}, the approximate deconvolution method(ADM) \cite{stolz1999,stolz2001,adams2004,layton2012,hickel2006,san2015,san2016}, the gradient model \cite{clark1979}, the dynamic Smagorinsky model (DSM) \cite{germano1991, germano1992, lilly1992}, the dynamic mixed models (DMM) \cite{zang1992,vreman1994,vreman1997,habisreutinger2007}, the Reynolds-stress-constrained large-eddy simulation model \cite{chen2012}, Implicit-LES (ILES) \cite{boris1992,garnier1999}, \emph{et al.}
  
Recently, machine learning approaches have been applied to turbulence modeling, including the Reynolds-Averaged Navier-Stokes (RANS) \cite{tracey2015, ling2015, ling2016, ling2016a, xiao2016, wang2017, wang2018a, wu2018, pan2018, pan2018a, zhu2019} and LES models \cite{sarghini2003, ma2019, srinivasan2019, wang2018, zhou2019, xie2019, xie2019a, xie2019b, xie2020a, xie2020b, xie2020c, rosofsky2020, gamahara2017, raissi2019, beck2019, yang2019, fukami2019, maulik2019, maulik2017, maulik2018a, maulik2019a, prat2020, pawar2020}. Ling \emph{et al.} proposed a machine learning strategy to predict the Reynolds stress tensor by embedding the Galilean invariance \cite{ling2015, ling2016, ling2016a}. Ma \emph{et al.} established a natural analogy between recurrent neural network and the Mori-Zwanzig formalism, and proposed a systematic approach for developing long-term reduced models to predict the Kuramoto-Sivashinsky equations and the Navier-Stokes equations \cite{ma2019}. The ANN models built with the filtered velocity gradients as input variables show good performance for both the \emph{a priori} and the \emph{a posteriori} studies in the prediction of isotropic turbulence \cite{wang2018, zhou2019, xie2019, xie2019a, xie2019b, xie2020a, xie2020b, xie2020c} and magneto-hydrodynamic turbulence \cite{rosofsky2020}, but show no advantage over the Smagorinsky model in the \emph{a posteriori} testing for the channel flow \cite{gamahara2017}. Raissi \emph{et al.} proposed a physical-informed neural network to learn the unclosed terms for turbulent scalar mixing \cite{raissi2019}. Beck \emph{et al.} employed the convolution neural network and residual neural network to obtain more accurate and stable LES models \cite{beck2019}. Besides, the artificial neural network also can be used for wall modeling in LES \cite{yang2019}, the turbulent inflow generator \cite{fukami2019}, and the classifier for turbulence modeling hypothesis \cite{maulik2019}, \emph{et al}. Data-driven blind deconvolution was proposed to recover the unfiltered turbulence fields without any \emph{a priori} knowledge \cite{maulik2017}. Maulik \emph{et al.} used two single-layer artificial neural networks to perform the convolution and deconvolution between coarse-grained unfiltered fields and filtered fields to model the SGS stress of decaying two-dimensional Kraichnan turbulence \cite{maulik2018a, maulik2019a}. A summary and systematic discussion of the recent developments on data-driven turbulence models can be referred to these reviews \cite{kutz2017, duraisamy2019}.    
  
Previously, we proposed a spatial artificial neural network (SANN) model \cite{xie2020a, xie2020b, xie2020c} for large eddy simulations of incompressible and compressible turbulence. It was shown that the flow dynamics at the scales between $\Delta$/2 and 2$\Delta$ are crucial for accurately reconstructing the SGS terms at the filter width $\Delta$ by using artificial neural networks. It was also found that numerical dissipation can be added individually to keep the stability of the SGS models without introducing considerable  numerical errors if the grid scale is smaller than the filter width. It is important to incorporating more \emph{a priori} knowledge of SGS stress to improve the efficiency of ANN-based SGS models, such as the use of deconvolution and convolution operators in constructing the SGS stress. 

In this paper, we propose a deconvolutional artificial neural network (DANN) framework for reconstructing the SGS stress based on DNS data of solenoidally forced stationary incompressible isotropic turbulence at grid resolution of $1024^{3}$. The main difference from the previous work is that we do not directly use the ANN to predict the SGS stress, but first employ the DANN framework to recover the unfiltered velocity with the stencil of the neighboring filtered velocities, and then use the recovered unfiltered velocity to reconstruct the SGS stress. The grid width of the DANN model is chosen to be smaller than the filter width, in order to accurately model the effect of SGS dynamics on the SGS energy flux. Comparative analysis between the DANN models and DNS are carried out in both \emph{a priori} and \emph{a posteriori} testings. The velocity spectrum and the SGS energy flux recovered by the proposed DANN models have been evaluated in the \emph{a priori} analysis. Compared to the conventional approximate deconvolution method (ADM) and velocity gradient model (VGM), the DANN models with reasonable stencils can predict the SGS stress more accurately, which makes the correlation coefficients larger than 99\% and the relative errors less than 15\%. In the \emph{a posteriori} study, we examine the filtered velocity spectrum, various statistics of the filtered velocity and the instantaneous coherent structures predicted by the DANN models. Compared with our SANN model in the previous work \cite{xie2020a}, the computational cost of the DANN model in LES is greatly reduced by nearly two orders of magnitude. Furthermore, the trained DANN models are applied to LES computations with different filter widths without any fine-tuning. This paper is organized as follows. Section \ref{sec:level2} briefly describes the governing equations and computational method. Section \ref{sec:level3} introduces DNS database of the incompressible turbulence. Section \ref{sec:level4} presents the deconvolutional artificial neural network models. Section \ref{sec:level5} provides both \emph{a priori} and \emph{a posteriori} results of the proposed DANN models. Conclusions are drawn in Section \ref{sec:level6}.

\section{Governing equations and numerical method} \label{sec:level2}
The Navier-Stokes equations in conservative form for incompressible turbulence are \cite{pope2000,sagaut2006}

 \begin{equation}
  \frac{{\partial {u_i}}}{{\partial {x_i}}} = 0,
  \label{ns1}
\end{equation}
 \begin{equation}
\frac{{\partial {u_i}}}{{\partial t}} + \frac{{\partial \left( {{u_i}{u_j}} \right)}}{{\partial {x_j}}} =  - \frac{{\partial p}}{{\partial {x_i}}} + \nu \frac{{{\partial ^2}{u_i}}}{{\partial {x_j}\partial {x_j}}} + {{\mathcal F}_i}, 
  \label{ns2}
\end{equation}
where ${{u}_{i}}$ is the i-th velocity component, p is the pressure, $\nu$ is the kinematic viscosity, and ${{\mathcal{F}}_{i}}$ is the i-th solenoidally large-scale force component. 

Besides, the total kinetic energy ${{E}_{k}}$, the Taylor microscale Reynolds number ${{\operatorname{Re}}_{\lambda }}$, the Kolmogorov length scale $\eta $, and the integral length scale ${{L}_{I}}$ are defined, respectively, as \cite{pope2000}
\begin{equation}
{E_k} = \frac{1}{2}\left\langle {{u_i}{u_i}} \right\rangle,
\label{Ek}
\end{equation}
\begin{equation}
  {{\operatorname{Re}}_{\lambda }}=\frac{{{u}^{rms}}\lambda }{\nu },
  \label{Re_l}
\end{equation}
\begin{equation}
  \eta ={{\left( \frac{{{\nu }^{3}}}{\epsilon } \right)}^{1/4}},
  \label{eta}
\end{equation}
\begin{equation}
  {L_I}=\frac{\pi}{2{{\left( {{u}^{rms}} \right)}^{2}}}\int_{0}^{+\infty }{\frac{E\left( k \right)}{k}dk}.
  \label{LI}
\end{equation}
Here, ${{u}^{rms}}=\sqrt{\left\langle {{u}_{i}}{{u}_{i}} \right\rangle /3}$ is the root mean square (rms) value of the velocity, where $\left\langle {} \right\rangle $ stands for a spatial average of the computational domain. $\nu$ is the kinematic viscosity. $\epsilon =2\nu \left\langle {{S}_{ij}}{{S}_{ij}} \right\rangle $ is the dissipation rate, where ${{S}_{ij}}=\frac{1}{2}\left( \partial {{u}_{i}}/\partial {{x}_{j}}+\partial {{u}_{j}}/\partial {{x}_{i}} \right)$ is the strain rate tensor. $\lambda ={{u}^{rms}}\sqrt{15\nu /\epsilon }$  is the Taylor microscale. ${E_k}=\int_{0}^{+\infty }{E\left( k \right)dk}$, where $E(k)$ is the spectrum of kinetic energy per unit mass.

\begin{table}[tbp]
	\begin{center}
		\normalsize
		\caption{ Parameters and statistics for DNS and fDNS at grid resolution of $1024^3$.}\label{tab:1}		
		\small
		\begin{tabular*}{0.95\textwidth}{@{\extracolsep{\fill}}lccccccccccc}
			\hline\hline
			${ {Re}_{\lambda }}$ & $E_{k}^{DNS}$ & $E_{k}^{fDNS}$ &  $\eta /{{\Delta }_{DNS}}$ & ${{L}_{I}}/\eta $ & $\lambda /\eta$ & ${{u}^{rms}}$ & ${{\omega}^{rms}}$ &  $\epsilon $ \\  \hline
			252   & 2.63 & 2.31 & 1.01  & 235.2 & 31.2 & 1.33 & 15.53 & 0.73 \\  \hline\hline	
		\end{tabular*}
		
	\end{center}
\end{table}

The physical quantities in turbulent flow can be separated into resolved large-scale and sub-filter small-scale quantities by introducing a spatial filtering operation \cite{lesieur1996, meneveau2000, leonard1975} $\bar{f}\left( \mathbf{x} \right)=\int\limits_{\Omega}{f\left( {\mathbf{{x}'}} \right)G\left( \mathbf{x}-\mathbf{{x}'}; \Delta  \right)d\mathbf{{x}'}}$, where an overbar represents a filtered variable, $\Omega$ is the overall domain, $G$ and $\Delta $ are the filter kernel and filter width, respectively. Therefore, the incompressible filtered Navier-Stokes equations for the resolved variables can be expressed as
 \begin{equation}
\frac{\partial {{{\bar{u}}}_{i}}}{\partial {{x}_{i}}}=0,
\label{fns1}
\end{equation}
\begin{equation}
\frac{\partial {{{\bar{u}}}_{i}}}{\partial t}+\frac{\partial \left( {{{\bar{u}}}_{i}}{{{\bar{u}}}_{j}} \right)}{\partial {{x}_{j}}}=-\frac{\partial \bar{p}}{\partial {{x}_{i}}}-\frac{\partial {{\tau }_{ij}}}{\partial {{x}_{j}}}+\nu\frac{{{\partial }^{2}}{{{\bar{u}}}_{i}}}{\partial {{x}_{j}}\partial {{x}_{j}}}+{{\bar{\mathcal{F}}}_{i}}.
\label{fns2}
\end{equation}
Here, the SGS stress tensor ${{\tau }_{ij}}$ at the right hand side of Eq. ~(\ref{fns2}) is defined by
\begin{equation}
{{\tau }_{ij}}=\overline{{{u}_{i}}{{u}_{j}}}-{{\bar{u}}_{i}}{{\bar{u}}_{j}}.
\label{tau}
\end{equation}
Obviously, the unclosed SGS stress can not be solved by the filtered governing equations directly.

In this paper, a pseudo-spectral method is applied to numerically simulate the incompressible homogeneous isotropic turbulence in a cubic box of ${{\left( 2\pi  \right)}^{3}}$ on a uniform grid with periodic boundary conditions. The aliasing errors are eliminated by the two-thirds rule \cite{patterson1971}. An explicit two-step Adams-Bashforth scheme with second-order time accuracy is employed for time marching \cite{wang2012}. The solenoidally large-scale forcing is constructed by fixing the total kinetic energy in the two lowest wavenumber shells \cite{wang2012a}.

\begin{figure}\centering
	\includegraphics[width=.7\textwidth]{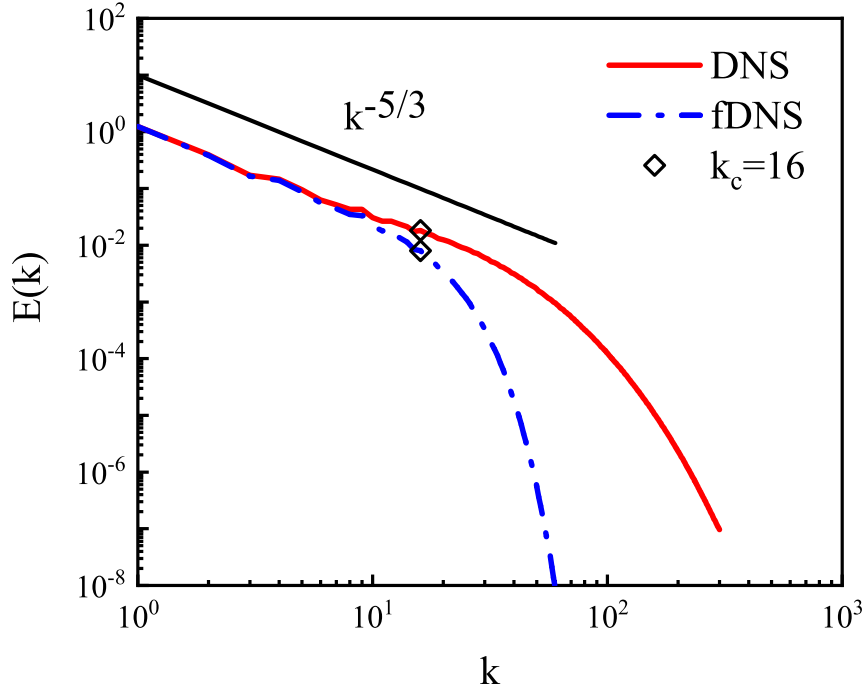}
	\caption{Velocity spectrum from DNS and filtered DNS of a forced incompressible isotropic turbulence. Diamond represents the cutoff wavenumber ${{k}_{c}}=16$ ($\Delta =32{{\Delta }_{DNS}}$).}\label{fig:1}
\end{figure}

\section{DNS database of incompressible turbulence} \label{sec:level3}
In this paper, the direct numerical simulation of a forced incompressible isotropic turbulence is performed with the grid resolution of $1024^{3}$ at Taylor Reynolds number ${{\operatorname{Re}}_{\lambda }}\approx 250$. The filtered physical quantities are calculated by the common Gaussian filter, which is calculated in one dimension by \cite{pope2000,sagaut2006,leonard1975}
\begin{equation}
G\left( x \right)={{\left( \frac{6}{\pi {{\Delta }^{2}}} \right)}^{1/2}}\exp \left( -\frac{6{{x}^{2}}}{{{\Delta }^{2}}} \right),
\label{G}
\end{equation}
where $\Delta =32{{\Delta }_{DNS}}$ is the filter width and ${{\Delta }_{DNS}}$ denotes the grid spacing of DNS. The corresponding cutoff wavenumber ${{k}_{c}}$ is ${{k}_{c}}=\pi / \Delta = 16$. The detailed one-point statistics for the incompressible turbulent flow are summarized in Table~\ref{tab:1} \cite{wang2018b,xie2020a}. The resolution parameter of DNS $\eta /{{\Delta }_{DNS}}$ is approximate to 1.01, and the corresponding resolution parameter ${{k}_{\max }}\eta \approx 2.11$, where the largest wavenumber $k_{max}=N/3$, and N is the number of grids in each direction. A resolution of ${{k}_{\max }}\eta \ge 2.1$ is sufficiently enough to achieve the convergence of the kinetic energy spectrum at different wavenumbers \cite{ishihara2007, ishihara2009}. 

The velocity spectrum of DNS and filtered DNS is shown in Fig.~\ref{fig:1}. The filtered velocity spectrum almost overlaps with the results of DNS at the inertial region with a ${{k}^{-5/3}}$ scaling, but decays steeply as the wavenumber becomes larger than the cutoff wavenumber. Nearly 88\% of the turbulent kinetic energy resides in the filtered flow field with the filter width $\Delta =32{{\Delta }_{DNS}}$.

The \emph{a priori} and \emph{a posteriori} analysis of subgrid modeling are presented in the paper. In the \emph{a priori} study of LES, data-driven deconvolution models are constructed by ANN. In the \emph{a posteriori} testing, the LES simulations modeled by the trained ANN are calculated for the grid resolution of ${{64}^{3}}\left( {{\Delta }_{LES}}=\Delta /2 \right)$ with the filter width $\Delta =32{{\Delta }_{DNS}}$ and corresponding cutoff wavenumber ${{k}_{c}}=16$, where ${{\Delta }_{LES}}$ represents the grid size of LES. 

\begin{figure}\centering
	\includegraphics[width=1.0\textwidth]{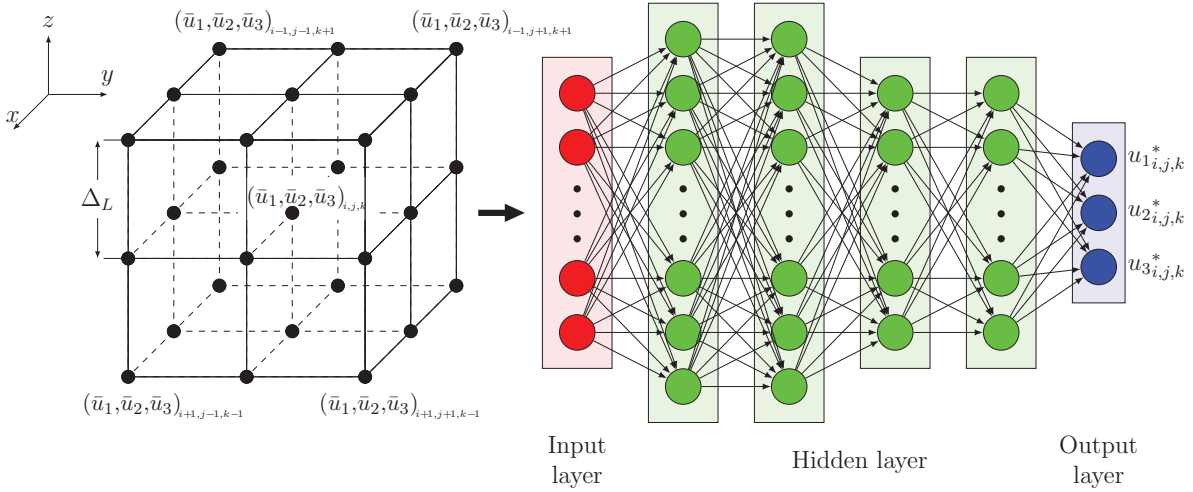}
	\caption{Schematic diagram of the deconvolutional artificial neural network structure.}\label{fig:2}
\end{figure}

\section{Deconvolutional neural networks} \label{sec:level4}
Our development of the DANN models is motivated by the approximate deconvolution method \cite{stolz1999,stolz2001,maulik2017,maulik2018a,maulik2019}. The original unfiltered variables can be approximately recovered by the deconvolution operation \cite{layton2012}. The SGS stress can be approximated as
\begin{equation}
{{\tau }_{ij}}=\overline{u_{i}^{*}u_{j}^{*}}-{{\bar{u}}_{i}}{{\bar{u}}_{j}},
\label{tau_ij_AD}
\end{equation}
where the asterisk denotes the approximately recovered variables.
The iterative Van Cittert procedure is usually used to reconstruct the unfiltered variables in ADM, namely \cite{sagaut2006,stolz2001}
\begin{equation}
u_{i}^{*}=A{{D}^{N}}\left( {{{\bar{u}}}_{i}} \right)=\sum\limits_{i=1}^{N}{{{\left( I-G \right)}^{i-1}}\left( {{{\bar{u}}}_{i}} \right)},
\label{ui_AD}
\end{equation}
where $A{{D}^{N}}$ is the abbreviation of the Nth approximate deconvolution. The SGS stress can be simply modeled by the Leonard stress for N=1, which is also called the scale-similarity model \cite{pope2000,sagaut2006}:
\begin{equation}
{{\tau }_{ij}}=\overline{{{{\bar{u}}}_{i}}{{{\bar{u}}}_{j}}}-{{\bar{\bar{u}}}_{i}}{{\bar{\bar{u}}}_{j}}.
\label{tau_sim}
\end{equation}
In order to reconstruct the unfiltered velocity from the filtered velocity, a six-layer fully connected ANN (input layer, four hidden layers, and output layer) is established as shown in Fig.~\ref{fig:2}. The ANN establishes the nonlinear mapping between the input and output variables. The ANN is composed of multiple layers which consist of many neurons. The neurons receive input signals from the previous layer, and transmit them to the next layer after the successive mathematical operation of linear weighted sum and the nonlinear activation by the transfer function. The transformation from layer l-1 to layer l is performed as \cite{maulik2017,maulik2018a}
\begin{equation}
s_{i}^{l}=\sum\limits_{j}{W_{ij}^{l}X_{j}^{l-1}},
\label{ANN1}
\end{equation}
\begin{equation}
X_{i}^{l}=\sigma \left( s_{i}^{l}+b_{i}^{l} \right), 
\label{ANN2}
\end{equation}
where $X_{i}^{l}$ represents the input signals from the ith neuron of the lth layer, $W_{ij}^{l}$ is the weight, $s_{i}^{l}$ is the weighted summation of all input signals from the (l-1)th layer, $b_{i}^{l}$ is the bias, and $\sigma $ is the nonlinear transfer function. 
The DANN models establish the relationship between the coarse-grained filtered velocities at the neighboring stencils with spacing width ${{\Delta }_{L}}$ and the associated unfiltered velocities as follows,
\begin{equation}
\begin{aligned}
 & \mathbb{M}:\left\{ {{{\bar{u}}}_{l\!+\!i,m\!+\!j,n\!+\!k}},{{{\bar{v}}}_{l\!+\!i,m\!+\!j,n\!+\!k}},{{{\bar{w}}}_{l\!+\!i,m\!+\!j,n+k}} \right\}\in {{\mathbb{R}}^{3\times {{D}^{3}}}}\to \left\{ u_{l,m,n}^{*},v_{l,m,n}^{*},w_{l,m,n}^{*} \right\}\in {{\mathbb{R}}^{3}}, \\ 
 & \left\{ i,j,k \right\}\in \left\{ -\left\lfloor D/2 \right\rfloor ,...,0,...,\left\lfloor D/2 \right\rfloor  \right\},
\end{aligned}
\end{equation}
where u, v and w are velocity components, D is the number of points in a direction of the stencil, the indices ${l}$, ${m}$ and ${n}$ denote the discrete spatial coordinate of the structured mesh, $\left\lfloor {} \right\rfloor $ is the round-down operation and the indices i, j and k refer the relative spatial indexing of the neighboring points with the spacing width ${{\Delta }_{L}}$. The DANN model for  $D^{3}$-point stencil with spacing width ${{\Delta }_{L}}=\Delta/n $ is abbreviated as DANN$(D,n)$.
The number of input variables for DANN$(D,n)$ is $3\times {{D}^{3}}$, and that of output variables is 3. The hyperparameters of the DANN models are determined by the grid search method and are listed in Table ~\ref{tab:2}. There are total six layers with neurons $3\times {{D}^{3}}:128:128:64:64:3$. The activation functions of the hidden layers and output layer are the Leaky-Relu function and linear function, respectively. The Leaky-Relu function is given by \cite{maas2013,xie2020a},
\begin{equation}
\label{lrelu}
\sigma \left( x \right)=\left( \begin{aligned}
& \ \ x,\ \ if\ x>0 \\ 
& \alpha x,\ \ if\ x\le 0 \\ 
\end{aligned} \right),\ \ where\ \alpha =0.2.
\end{equation}

Mean square error (MSE) criterion is chosen as the loss function of DANN, which is defined as $L=\frac{1}{3}\left\langle \sum\limits_{i=1}^{3}{{{({u}_{i}-u_{i}^{*})}^{2}}} \right\rangle $, where $\hat{u}_{i}^{*}$ represents the predicted values of DANN, and $\left\langle {} \right\rangle $ denotes the average of the entire datasets. We adopt the cross-validation strategy to suppress parameter overfitting and divide the dataset into training dataset and testing dataset. We evaluate the general performance of the model for the testing dataset after completing a round of training. $2\times {{64}^{3}}$ samples are chosen from a snapshot of the DNS data with the degrees of freedom being ${{1024}^{3}}$. 70\% of samples are used to generate the training dataset and the others are for testing. The weights of DANN are optimized by the Adam algorithm \cite{kingma2017} for 2000 iterations with batch size and learning rate being 1024 and 0.01, respectively. The hyperparameters of DANN (the number of layers and neurons, etc.) are determined by the grid search method \cite{maulik2018a, maulik2019a}. We use GPU cores (NVIDIA GeForce RTX 2080Ti) to accelerate the training of the DANN models. Each DANN model is trained with a single GPU and summaries of the GPU training time for 2000 epochs with different stencil sizes and spacing widths are listed in Table~\ref{tab:3}. As the stencil size increases, the number of input variables has a dramatic increase. However, the number of total hyperparameters does not increase significantly with the number of input variables, therefore GPU training time increases slightly, and overall the training time is not sensitive to the stencil size.

In order to improve the robustness of the DANN model, we normalize all input and output variables to zero mean and unit variance by standard scaling before training. Due to the lack of standard derivation of unsolved variables in the \emph{a posterior} study, both input and output variables are rescaled by the mean and standard deviation of filtered variables, which are given by, respectively \cite{maulik2017, maulik2018a, maulik2019a}
\begin{equation}
{{{\bar{u}}'}_{i}}=\frac{{{{\bar{u}}}_{i}}-\left\langle {{{\bar{u}}}_{i}} \right\rangle }{std\left( {{{\bar{u}}}_{i}} \right)},
\end{equation}
\begin{equation}
u{{_{i}^{*}}^{\prime }}=\frac{u_{i}^{*}-\left\langle {{{\bar{u}}}_{i}} \right\rangle }{std\left( {{{\bar{u}}}_{i}} \right)},
\end{equation}
where $std\left( {{{\bar{u}}}_{i}} \right)=\sqrt{\left\langle {{\left( {{{\bar{u}}}_{i}}-\left\langle {{{\bar{u}}}_{i}} \right\rangle  \right)}^{2}} \right\rangle }$ denotes the standard deviation of ${{\bar{u}}_{i}}$.

\begin{table}[tbp]
	\begin{center}
		\caption{ Parameters of the DANN model.}\label{tab:2}
		\small
		\begin{tabular*}{0.95\textwidth}{@{\extracolsep{\fill}}lccccc}
			\hline\hline
			Layer structure  & Dataset & Training/Testing & Epoch & Batch size & Learning rate \\  \hline
			$3 \!\times\! D^{3}\!:\!128\!:\!128\!:\!64\!:\!64\!:\!3$ & $2 \!\times\! 64^{3}$ & 0.7/0.3 & 2000  & 1024  & 0.01 \\
			\hline\hline
		\end{tabular*}
		
	\end{center}
\end{table}%

\begin{table}[tbp]
	\begin{center}
		\caption{ The number of input variables and the GPU execution time for 2000-epoch training of the DANN(D,n) models with $D^3$-point stencil and spacing width ${{\Delta }_{L}}=n\Delta $.}\label{tab:3}
		\begin{tabular*}{0.95\textwidth}{@{\extracolsep{\fill}} lcccc }
			\hline\hline
			\small    
			DANN(D,n) & DANN(3,1) & DANN(5,1) & DANN(5,2) & DANN(9,2) \\ \hline
			Number of Input & 81    & 375   & 375   & 2187 \\
			Training time (GPU$\cdot$h) & 2.1     & 2.2   & 2.2   & 2.3 \\ \hline\hline
		\end{tabular*}%
	\end{center}
\end{table}

\begin{figure}\centering
	\begin{subfigure}{0.5\textwidth}
		\centering
		\includegraphics[width=0.9\linewidth]{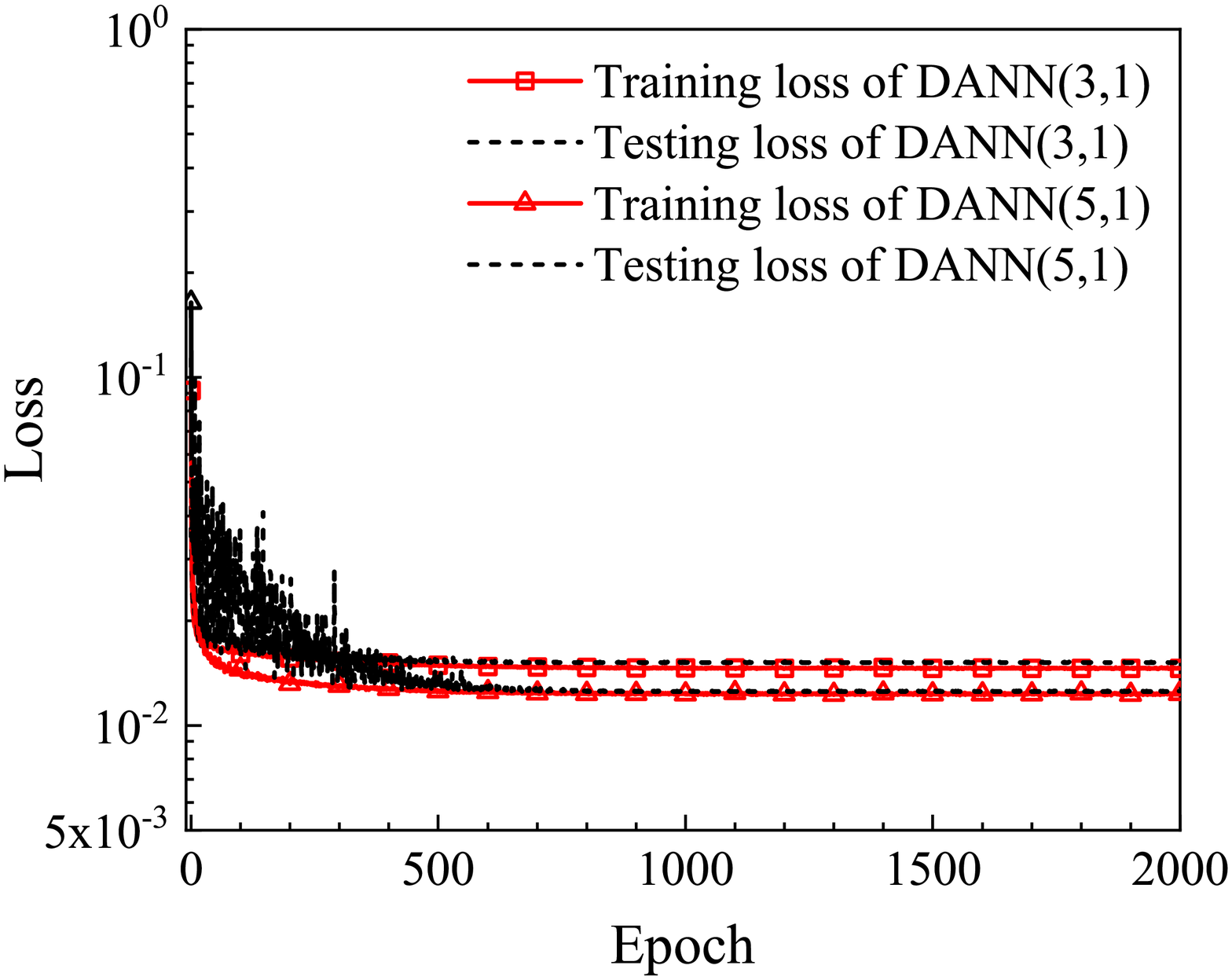}
		\caption{$\Delta/{{\Delta }_{L}}=1$}
	\end{subfigure}%
	\begin{subfigure}{0.5\textwidth}
		\centering
		\includegraphics[width=0.9\linewidth]{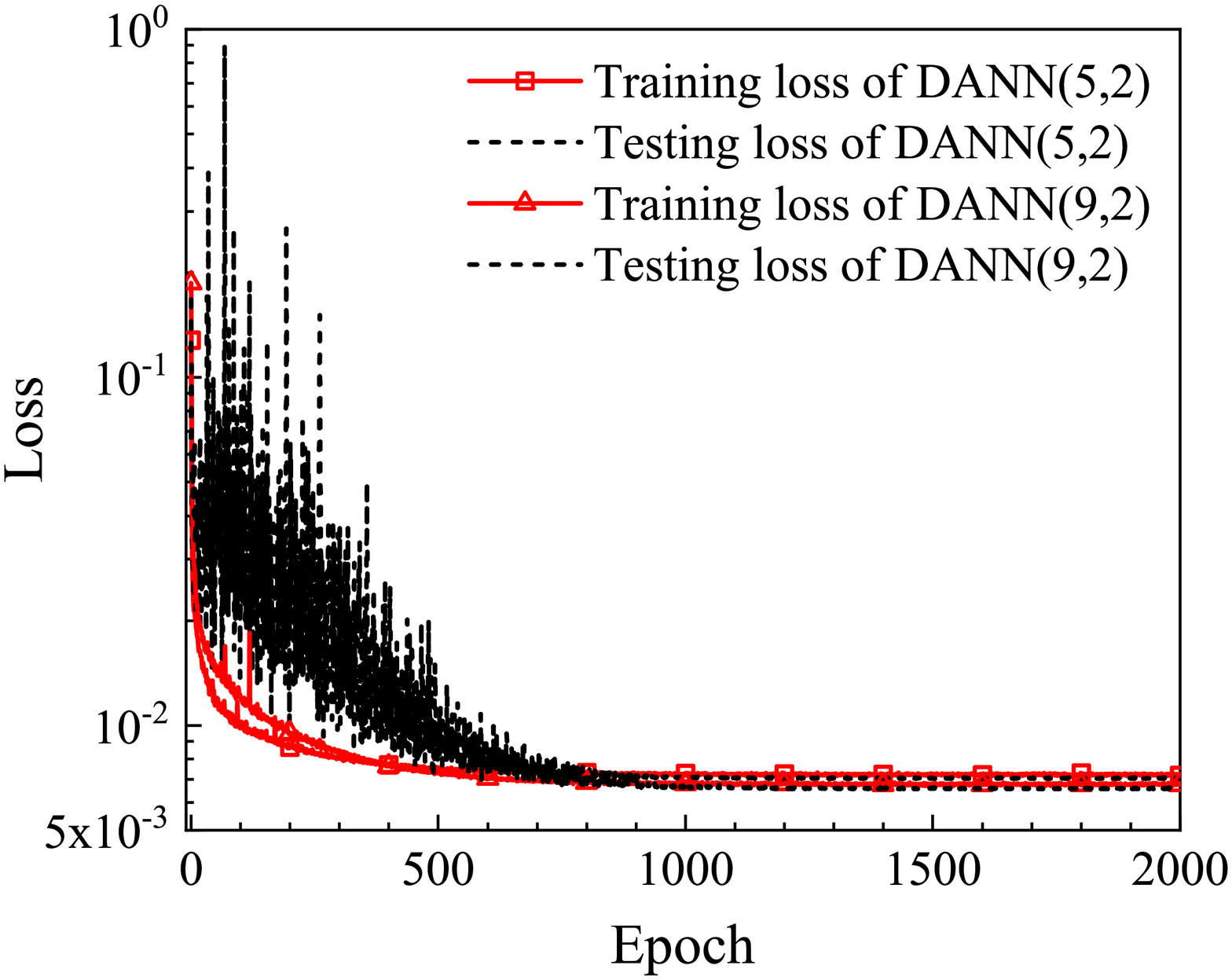}
		\caption{$\Delta/{{\Delta }_{L}}=2$}
	\end{subfigure}%
	
	\caption{ Learning curves of the proposed DANN model: (a) $\Delta$/$\Delta_L$=1, (b) $\Delta$/$\Delta_L$=2.}\label{fig:3}
\end{figure}

The learning curves of the DANN models for different stencil sizes and spacing widths are shown in Fig.~\ref{fig:3}. After a long training for the DANN models with 2000 epochs, the MSE losses in both training datasets and testing datasets are gradually converging and become stationary. The testing loss is very close to the training loss for all cases, which means that the hyperparameter selection is suitable and the DANN models are well-trained and not overfitting.

\section{results of the DANN models} \label{sec:level5}
In this section, we evaluate the performance of our proposed DANN models for the incompressible forced isotropic turbulence at ${{Re}_{\lambda }}\approx 250$ in both \emph{a priori} and \emph{a posteriori} studies. In the \emph{a priori} study, the DANN models are tested by predicting a snapshot of DNS data which is different from the dataset for the DANN training via three assessment indicators: correlation coefficient ($C$), the relative error ($E_{r}$), and the root-mean-square value ($R$). In the \emph{a posterior} study, the proposed DANN models can predict the statistics and instantaneous structures of turbulence with higher accuracy than the ILES, DSM and DMM models at grid resolution of $64^3$ ($\Delta_{LES}=\Delta/2$). Furthermore, we have performed an \emph{a posteriori} study of the DANN models for LES at different filter widths ($\Delta=16\Delta_{DNS}$ and $64\Delta_{DNS}$). The results indicate that the proposed DANN models have a strong generalization capability. 

\subsection{\emph{A priori} study}
In order to evaluate the performance of different models, three metrics are used to measure the difference between the predicted value ($Q^{model}$) and the true value ($Q^{real} $) for any targeted physical quantity $Q$ (i.e. the unsolved velocity $u_i^{*}$ and the SGS stress $\tau_{ij}$, etc.). They are the correlation coefficient $C\left( Q \right)$, the relative error ${{E}_{r}}\left( Q \right)$, and the root-mean-square value $R\left( Q \right)$, which are defined, respectively, by \cite{gamahara2017,xie2019,xie2019a,xie2019b,xie2020a,xie2020b,xie2020c,rosofsky2020}

\begin{equation}
C\left( Q \right) = \frac{{\left\langle {\left( {{Q^{real}} - \left\langle {{Q^{real}}} \right\rangle } \right)\left( {{Q^{model}} - \left\langle {{Q^{model}}} \right\rangle } \right)} \right\rangle }}{{{{\left\langle {{{\left( {{Q^{real}} - \left\langle {{Q^{real}}} \right\rangle } \right)}^2}} \right\rangle }^{1/2}}{{\left\langle {{{\left( {{Q^{model}} - \left\langle {{Q^{model}}} \right\rangle } \right)}^2}} \right\rangle }^{1/2}}}},
\end{equation}

\begin{equation}
{E_r}\left( Q \right) = \frac{{{{\left\langle {{{\left( {{Q^{real}} - {Q^{model}}} \right)}^2}} \right\rangle }^{1/2}}}}{{{{\left\langle {{{\left( {{Q^{real}}} \right)}^2}} \right\rangle }^{1/2}}}},
\end{equation}

\begin{equation}
R\left( Q \right) = {\left\langle {{{\left( {Q - \left\langle Q \right\rangle } \right)}^2}} \right\rangle ^{1/2}},
\end{equation}
where $\left\langle \centerdot  \right\rangle $ denotes spatial average of the entire domain. Results with the high correlation coefficient and low relative error indicate the success of modeling. The classical approximate deconvolution models $AD^N$ (N$\le $5) and the velocity gradient model are used to compare with the DANN models. The $AD^N$ models are calculated by the iterative Van-Cittert algorithm in Eq.~\ref{ui_AD}. The velocity gradient model is defined as \cite{clark1979, zang1992, vreman1994, vreman1997}, 
\begin{equation}
{\tau _{ij}} = \frac{{{\Delta ^2}}}{{12}}\frac{{\partial {{\bar u}_i}}}{{\partial {x_k}}}\frac{{\partial {{\bar u}_j}}}{{\partial {x_k}}}.
\label{tau_vgm}
\end{equation}

Table~\ref{tab:4} shows the correlation coefficients, relative errors, and root-mean-square values of the approximate unfiltered velocity component $u_1^*$ in both training and testing datasets after the 2000-iteration training process. The results in the other two velocity components ($u_2^*$ and $u_3^*$) are similar to $u_1^*$. The slight difference between the results of these two datasets shows that all trained DANN models with different stencil sizes and spacing widths are not overfitting. 

\begin{table}[tbp]
	\begin{center}
		\caption{Correlation coefficient ($C$), relative error ($E_r$), and root-mean-square value ($R$) of $u_1^*$ in different datasets for different DANN models $\Delta/{{\Delta }_{L}}=n $.} 	\label{tab:4}%
		\begin{tabular*}{0.95\textwidth}{@{\extracolsep{\fill}} lcccc }
			\hline\hline
			\small   
			Dataset\textbackslash{}$C(u_1^*)$ & DANN(3,1) & DANN(5,1) & DANN$(5,2)$ & DANN$(9,2)$ \\  \hline
			Training & 0.988 & 0.989 & 0.995 & 0.995 \\ 
			Testing  & 0.988 & 0.988 & 0.995 & 0.995 \\  \hline
			Dataset\textbackslash{}$E_r(u_1^*)$ & DANN(3,1) & DANN(5,1) & DANN$(5,2)$ & DANN$(9,2)$ \\  \hline
			Training & 0.152 & 0.152 & 0.105 & 0.101 \\
			Testing  & 0.155 & 0.152 & 0.104 & 0.101 \\  \hline
			Dataset\textbackslash{}$R(u_1^*)$ & DANN(3,1) & DANN(5,1) & DANN$(5,2)$ & DANN$(9,2)$ \\  \hline
			Training & 1.516 & 1.510 & 1.529 & 1.522 \\
			Testing  & 1.518 & 1.513 & 1.529 & 1.522 \\ \hline\hline
		\end{tabular*}%
	\end{center}
\end{table}%

\begin{table}[tbp]
	\begin{center}
		\caption{Comparisons of the correlation coefficients ($C$), relative errors ($E_r$), and root-mean-square values ($R$) of $u_i^*$ for different DANN models and classical $AD^N$ models.} 
		\label{tab:5}%
		\begin{tabular*}{0.95\textwidth}{@{\extracolsep{\fill}} 
				p{0.8cm}<{\centering}p{1.5cm}<{\centering}p{1.5cm}<{\centering}p{1.5cm}<{\centering}p{1.5cm}<{\centering}p{0.8cm}<{\centering}p{0.8cm}<{\centering}p{0.8cm}<{\centering}p{0.8cm}<{\centering}p{0.8cm}<{\centering} }
			\hline\hline
			\small 		
			$C(u_i^*)$ & DANN(3,1) & DANN(5,1) & DANN(5,2) & DANN(9,2) & $AD^1$ & $AD^2$ & $AD^3$ & $AD^4$ & $AD^5$  \\ \hline
			$u_1^*$  & 0.989 & 0.990 & 0.995 & 0.995 & 0.982 & 0.988 & 0.991 & 0.992 & 0.993 \\
			$u_2^*$  & 0.982 & 0.983 & 0.991 & 0.992 & 0.969 & 0.979 & 0.983 & 0.986 & 0.987 \\
			$u_3^*$  & 0.985 & 0.985 & 0.993 & 0.993 & 0.974 & 0.983 & 0.986 & 0.988 & 0.989 \\ \hline
			$E_r(u_i^*)$ & DANN(3,1) & DANN(5,1) & DANN(5,2) & DANN(9,2) & $AD^1$ & $AD^2$ & $AD^3$ & $AD^4$ & $AD^5$  \\ \hline
			$u_1^*$  & 0.153 & 0.151 & 0.104 & 0.101 & 0.208 & 0.163 & 0.145 & 0.134 & 0.127 \\    
			$u_2^*$  & 0.200 & 0.196 & 0.137 & 0.134 & 0.287 & 0.221 & 0.196 & 0.181 & 0.171 \\
			$u_3^*$  & 0.184 & 0.180 & 0.125 & 0.122 & 0.255 & 0.199 & 0.176 & 0.163 & 0.155 \\ \hline
			$R(u_i^*)$  & DANN(3,1) & DANN(5,1) & DANN(5,2) & DANN(9,2) & DNS & $AD^2$ & $AD^3$ & $AD^4$ & $AD^5$  \\ \hline 
			$u_1^*$  & 1.507 & 1.503 & 1.521 & 1.515 & 1.527 & 1.486 & 1.496 & 1.501 & 1.504 \\      
			$u_2^*$  & 1.123 & 1.122 & 1.135 & 1.136 & 1.146 & 1.091 & 1.105 & 1.112 & 1.116 \\
			$u_3^*$  & 1.239 & 1.242 & 1.253 & 1.252 & 1.262 & 1.213 & 1.225 & 1.231 & 1.235 \\ \hline\hline
		\end{tabular*}%
	\end{center}
\end{table}%

The comparisons of correlation coefficients, relative errors, and root-mean-square values of the approximate deconvolved velocity $u_i^*$ for different models are summarized in Table~\ref{tab:5}. The correlation coefficients $C(u_1^*)$ and relative errors $E_r(u_1^*)$ of the DANN(3,1) model are respectively approximate to 0.989 and 0.153, which means that the performance of the DANN(3,1) model is very close to the $AD^2$ method. With the increasing of stencil size, more spatial information is included in the DANN models, which result in higher correlation coefficients and lower relative errors. As the spacing width becomes a half of the filter width, the DANN models perform much better than the classical $AD^N$ (N$\le $5) methods and the root-mean-square values are very close to the DNS data. For the DANN(5,2) model, the correlation coefficient and relative error are $C(u_1^*) \approx 99.5\%$ and $E_r(u_1^*) \approx 10.4\%$, respectively. The root-mean-square value of the DANN(5,2) model is equal to 1.521, which is very close to that of the DNS data ($R_{DNS}(u_1^*)\approx 1.527$). 

\begin{figure}\centering
	\begin{subfigure}{0.5\textwidth}
		\centering
		\includegraphics[width=0.9\linewidth]{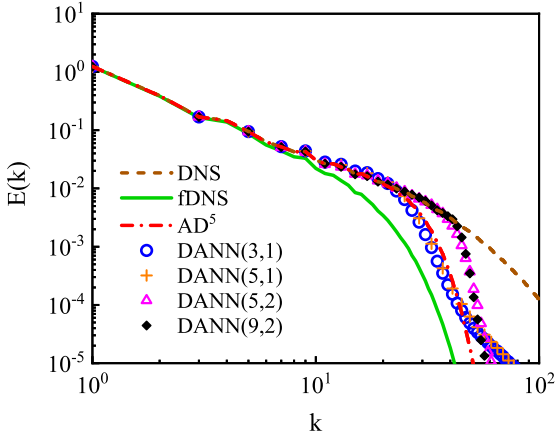}
		\caption{N=1024 and k $\le$ 100}
	\end{subfigure}%
	\begin{subfigure}{0.5\textwidth}
		\centering
		\includegraphics[width=0.9\linewidth]{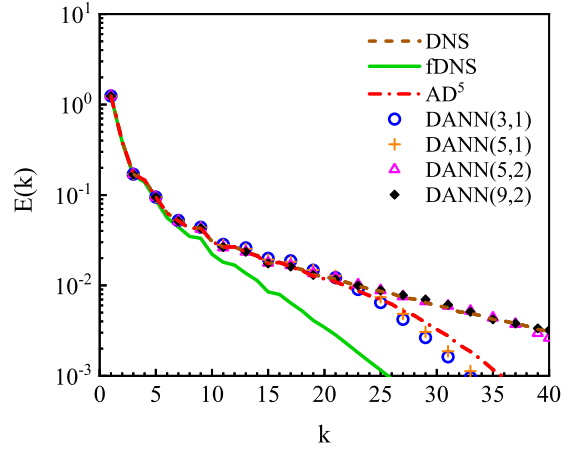}
		\caption{N=1024 and k $\le$ 40}
	\end{subfigure}%
	
	\caption{Comparisons of velocity spectrum for different SGS models in the \emph{a priori} analysis.}\label{fig:4}
\end{figure}

The spectra of approximate unfiltered velocity for different DANN models are shown in Fig.~\ref{fig:4}. The unfiltered velocity spectra given by the DANN(3,1) and DANN (5,1) models are similar to that of $AD^5$. The spectra predicted by the DANN(5,2) and the DANN (9,2) models are closer to that of DNS at high wavenumbers ($30 \le k \le 50$), indicating that the DANN models perform better than the classical ADM models in terms of ability to recover unfiltered velocity. 

In this paper, we consider two forms of the DANN models: the direct modeling ($\tau_{ij}^{DANN-D}$), and the scale-similarity form ($\tau_{ij}^{DANN}$), which are respectively written as \cite{bardina1980, liu1994},
\begin{equation}
{{\tau }_{ij}^{DANN-D}}=\overline{u_{i}^{*}u_{j}^{*}}-{{\bar{u}}_{i}}{{\bar{u}}_{j}},
\label{tau_ij_d}
\end{equation}
\begin{equation}
{{\tau }_{ij}^{DANN}}=\overline{u_{i}^{*}u_{j}^{*}}-{{\bar{u_{i}^{*}}}}{{\bar{u_{j}^{*}}}}.
\label{tau_ij_s}
\end{equation}

Tables~\ref{tab:6}$-$~\ref{tab:8} show correlation coefficients, relative errors and root-mean-square values of $\tau_{ij}$ for direct SGS models (DANN-D) and scale-similarity SGS models (DANN). The DANN models perform better than the DANN-D models with higher correlation coefficients and lower relative errors. When the ratio of spacing width and the number of neighboring points increase, the performances of the DANN models improve significantly and become much better than the $AD^N$ models. The correlation coefficients of the DANN(9,2) model reach 99\% and the relative errors are less than 15\%. In comparison, the relative errors of the $AD^5$ model are more than 35\%. The root-mean-square values predicted by DANN models are also very close to that of DNS data. 

\begin{table}[tbp]
	\begin{center}
		\caption{Comparisons of the correlation coefficients of $\tau_{ij}$ for different SGS models (VGM, $AD^{N}$, DANN-D, and DANN) in the \emph{a priori} analysis.} 
		\label{tab:6}%
		\begin{tabular*}{0.95\textwidth}{@{\extracolsep{\fill}} lcccccc}
			\hline\hline
			\small 	
			Model\textbackslash{}C($\tau_{ij}$) & $\tau_{11}$ & $\tau_{22}$ & $\tau_{33}$ & $\tau_{12}$ & $\tau_{23}$ & $\tau_{13}$ \\ \hline
			VGM      & 0.915 & 0.893 & 0.900 & 0.918 & 0.929 & 0.926 \\			
			$AD^1$   & 0.161 & 0.150 & 0.134 & 0.144 & 0.152 & 0.158 \\
			$AD^2$   & 0.607 & 0.632 & 0.599 & 0.557 & 0.556 & 0.560 \\
			$AD^3$   & 0.818 & 0.831 & 0.813 & 0.788 & 0.784 & 0.789 \\
			$AD^4$   & 0.896 & 0.900 & 0.891 & 0.883 & 0.880 & 0.884 \\
			$AD^5$   & 0.930 & 0.931 & 0.926 & 0.925 & 0.924 & 0.926 \\
			DANN(3,1)-D & 0.748 & 0.813 & 0.767 & 0.729 & 0.708 & 0.719 \\
			DANN(5,1)-D & 0.776 & 0.839 & 0.804 & 0.754 & 0.730 & 0.747 \\
			DANN(5,2)-D & 0.956 & 0.964 & 0.959 & 0.950 & 0.946 & 0.949 \\
			DANN(9,2)-D & 0.969 & 0.967 & 0.967 & 0.958 & 0.960 & 0.958 \\
			DANN(3,1) & 0.945 & 0.947 & 0.940 & 0.946 & 0.948 & 0.950 \\
			DANN(5,1) & 0.950 & 0.954 & 0.948 & 0.951 & 0.952 & 0.954 \\
			DANN(5,2) & 0.988 & 0.987 & 0.987 & 0.988 & 0.990 & 0.990 \\
			DANN(9,2) & 0.990 & 0.989 & 0.989 & 0.991 & 0.992 & 0.992 \\
 			\hline\hline
		\end{tabular*}%
	\end{center}
\end{table}%

\begin{table}[tbp]
	\begin{center}
		\caption{Comparisons of the relative errors of $\tau_{ij}$ for different SGS models (VGM, $AD^{N}$, DANN-D, and DANN) in the \emph{a priori} analysis.} 		
		\label{tab:7}%
		\begin{tabular*}{0.95\textwidth}{@{\extracolsep{\fill}} lcccccc}
			\hline\hline
			\small 	
			Model\textbackslash{}$E_r(\tau_{ij}$) & $\tau_{11}$ & $\tau_{22}$ & $\tau_{33}$ & $\tau_{12}$ & $\tau_{23}$ & $\tau_{13}$ \\ \hline
			VGM       & 0.493 & 0.524 & 0.523 & 0.427 & 0.400 & 0.408 \\
			$AD^1$    & 1.329 & 1.209 & 1.276 & 1.924 & 2.003 & 1.968 \\
			$AD^2$    & 0.690 & 0.666 & 0.691 & 0.946 & 0.973 & 0.958 \\
			$AD^3$    & 0.493 & 0.490 & 0.503 & 0.633 & 0.646 & 0.636 \\
			$AD^4$    & 0.397 & 0.401 & 0.410 & 0.483 & 0.489 & 0.482 \\
			$AD^5$    & 0.340 & 0.347 & 0.354 & 0.396 & 0.397 & 0.392 \\
			DANN(3,1)-D & 0.521 & 0.437 & 0.491 & 0.818 & 0.874 & 0.852 \\
			DANN(5,1)-D & 0.517 & 0.414 & 0.455 & 0.785 & 0.834 & 0.793 \\
			DANN(5,2)-D & 0.204 & 0.194 & 0.202 & 0.335 & 0.351 & 0.342 \\
			DANN(9,2)-D & 0.209 & 0.186 & 0.191 & 0.308 & 0.302 & 0.311 \\
			DANN(3,1) & 0.315 & 0.303 & 0.324 & 0.337 & 0.332 & 0.326 \\  
			DANN(5,1) & 0.302 & 0.284 & 0.298 & 0.321 & 0.318 & 0.315 \\  
			DANN(5,2) & 0.136 & 0.141 & 0.142 & 0.157 & 0.149 & 0.150 \\  
			DANN(9,2) & 0.129 & 0.132 & 0.135 & 0.142 & 0.134 & 0.135 \\	
			\hline\hline
		\end{tabular*}%
	\end{center}
\end{table}%

\begin{table}[tbp]
	\begin{center}
		\caption{Comparisons of the root-mean-square values of $\tau_{ij}$ for different SGS models (VGM, $AD^{N}$, DANN-D, and DANN) in the \emph{a priori} analysis.} 			
		\label{tab:8}%
		\begin{tabular*}{0.95\textwidth}{@{\extracolsep{\fill}} lcccccc}
			\hline\hline
			\small 	
			Model\textbackslash{}$R(\tau_{ij}$) & $\tau_{11}$ & $\tau_{22}$ & $\tau_{33}$ & $\tau_{12}$ & $\tau_{23}$ & $\tau_{13}$ \\ \hline
			DNS    & 0.201 & 0.184 & 0.183 & 0.095 & 0.102 & 0.102 \\
			VGM    & 0.143 & 0.124 & 0.123 & 0.075 & 0.082 & 0.082 \\
			$AD^1$ & 0.302 & 0.225 & 0.250 & 0.171 & 0.193 & 0.189  \\ 
			$AD^2$ & 0.173 & 0.137 & 0.146 & 0.095 & 0.107 & 0.104  \\ 
			$AD^3$ & 0.158 & 0.133 & 0.136 & 0.083 & 0.093 & 0.091  \\ 
			$AD^4$ & 0.159 & 0.137 & 0.138 & 0.082 & 0.091 & 0.090  \\ 
			$AD^5$ & 0.162 & 0.142 & 0.142 & 0.083 & 0.091 & 0.090  \\ 
			DANN(3,1)-D & 0.205 & 0.179 & 0.180 & 0.110 & 0.123 & 0.121  \\
			DANN(5,1)-D & 0.203 & 0.180 & 0.186 & 0.110 & 0.121 & 0.117  \\
			DANN(5,2)-D & 0.194 & 0.178 & 0.176 & 0.098 & 0.106 & 0.105  \\
			DANN(9,2)-D & 0.192 & 0.176 & 0.173 & 0.097 & 0.104 & 0.104  \\
			DANN(3,1) & 0.163 & 0.153 & 0.146 & 0.086 & 0.092 & 0.092 \\
			DANN(5,1) & 0.164 & 0.156 & 0.152 & 0.087 & 0.093 & 0.092 \\
			DANN(5,2) & 0.187 & 0.172 & 0.170 & 0.094 & 0.101 & 0.100 \\
			DANN(9,2) & 0.186 & 0.171 & 0.168 & 0.094 & 0.100 & 0.100 \\
			\hline\hline
		\end{tabular*}%
	\end{center}
\end{table}%

\begin{figure}\centering
	\includegraphics[width=0.7\linewidth]{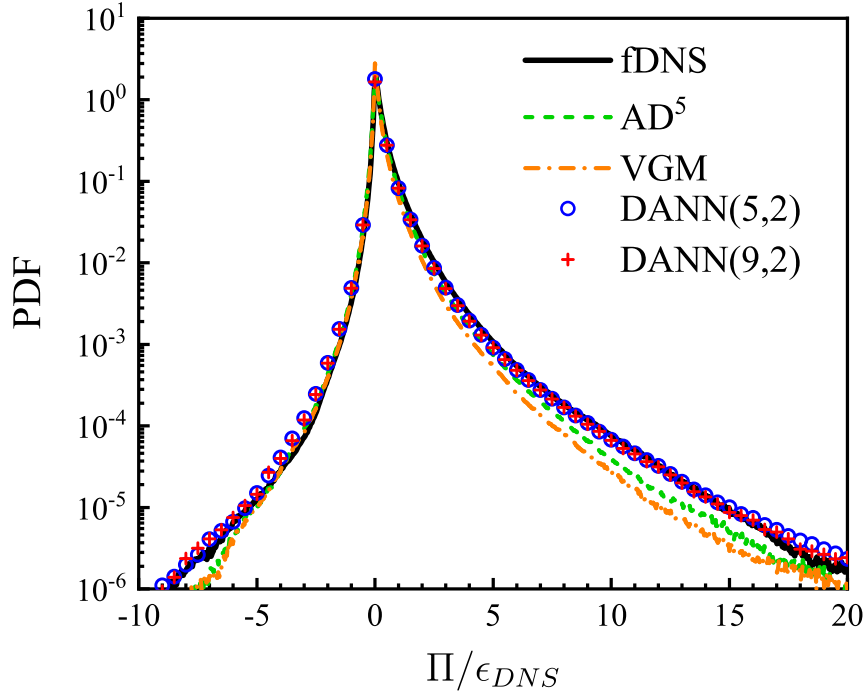}
	\caption{ PDFs of the normalized SGS flux $\Pi/\epsilon_{DNS}$ for scale-similarity SGS models with DANN (DANN) in emph{a priori} study.}\label{fig:5}
\end{figure}

Furthermore, we evaluate the performance of the SGS models on the SGS flux of kinetic energy \cite{sagaut2018, lesieur1996, meneveau2000}:
\begin{equation}
\Pi=-\tau_{ij}\bar{S}_{ij},\quad \bar{S}_{ij}=\frac{1}{2}(\frac{\partial{\bar{u}_i}}{\partial{x_j}}+\frac{\partial{\bar{u}_j}}{\partial{x_i}}),
\label{SGS_flux}
\end{equation}
where $\bar{S}_{ij}$ is the filtered strain-rate tensor. The SGS energy flux is normalized by the energy dissipation rate $\epsilon_{DNS}$ calculated from the DNS data. The probability density functions (PDFs) of the normalized SGS flux $\Pi/\epsilon_{DNS}$ with different stencil spacing sizes of DANN models ($\Delta/\Delta_L$=2) are shown in Fig.~\ref{fig:5}. The PDFs of the SGS energy flux with DANN(5,2) and DANN(9,2) are very close to the filtered DNS data. Compared to the $AD^5$ model and the VGM model, the positive PDF tail of the SGS flux can be predicted more accurately by DANN models. 

\subsection{\emph{A posteriori} study}
It is crucial to evaluate the performances of the SGS models in an \emph{a posteriori} tests, which are affected by both errors of SGS models and numerical schemes \cite{lesieur1996, meneveau2000}. We compared the DANN models with the classical LES models, including the implicit large eddy simulation (ILES), the dynamic Smagorinsky model (DSM), and the dynamic mixed model (DMM) \cite{xie2020a, xie2020b, xie2020c, xie2020}. For ILES, numerical dissipation is required to maintain the stability of numerical simulation of turbulence at the coarse grid. In this paper, we use a dissipative explicit hyper-viscosity scheme with fourth-order accuracy \cite{xie2020a}: 
\begin{equation}
\bar{u}_i^{n+1}=\bar{u}_i^n-\Delta t \cdot C^I_0(\frac{k}{k_0})^m \bar{u}_i^n, \quad k_0=\frac{2\pi}{3\Delta_{LES}}=\frac{4\pi}{3\Delta},
\label{iles}
\end{equation} 
where m=4, $C^I_0$=3, and $k_0$ is the largest wavenumber of LES truncated by the two-thirds dealiasing rule. The same treatment of numerical viscosity is also applied to the DSM, DMM, and DANN models. The average computational cost for different LES models at resolution of $64^3$ ($\Delta_{LES}=\Delta/2$) with the same filter width $\Delta=32\Delta_{DNS}$ are listed in Tables~\ref{tab:9}. For the DANN models, with the increasing of stencil size, computation time increases significantly since the number of input features grows sharply with the cube of stencil size. The average computation time of the DANN(5,2) model is approximately 1.3 times that of the DMM model. Compared with our SANN model in the previous work \cite{xie2020a}, the computational cost of the DANN model in LES is greatly reduced by nearly two orders of magnitude. In order to balance the accuracy and efficiency of the DANN models, we select DANN(5,2) with $5^3$-point stencil and spacing width $\Delta/{{\Delta}_{L}}=2$ as the representation of the DANN models for further discussions. The ratio of the time steps for LES and DNS is $\Delta t_{LES}/\Delta t_{DNS}=8$.

\begin{table}[tbp]
	\begin{center}
		\caption{The average computational cost of SGS stress $\tau_{ij}$ for different LES models at resolution of $64^3$ ($\Delta_{LES}=\Delta/2$) with the same filter width $\Delta=32\Delta_{DNS}$.} 			
		\label{tab:9}%
		\begin{tabular*}{0.95\textwidth}{@{\extracolsep{\fill}} lcccccc}
			\hline\hline
			\small 	
			Model       & DSM   & DMM    & DANN(3,1) & DANN(5,1) & DANN(5,2) & DANN(9,2) \\ \hline
			t(CPU$\cdot$s) & 0.895 & 1.694  & 1.164 & 2.128 & 2.130 & 7.760 \\
			t/t$_{DMM}$ & 0.528 & 1.000  & 0.687 & 1.256 & 1.257 & 4.581 \\
			\hline\hline
		\end{tabular*}%
	\end{center}
\end{table}%

The Smagorinsky model is an eddy-viscosity model which establishes the relationship between the SGS stress and the filtered strain rate tensor. The dynamic Smagorinsky model (DSM) is based on the Germano identity, which dynamically determine the model coefficients by the least squares method. The deviatoric part of the SGS stress is modeled by \cite{smagorinsky1963},
\begin{equation}
\tau_{ij}-\frac{\delta_{ij}}{3}\tau_{kk}=-2C_S^2{\Delta}^2|\bar{S}|\bar{S}_{ij},
\label{dsm}
\end{equation} 
where $|\bar{S}|=(2\bar{S}_{ij}\bar{S}_{ij})^{1/2}$ is the characteristic filtered strain rate. The model coefficient of the DSM model can be calculated dynamically by \cite{germano1991},
\begin{equation}
C_S^2=\frac{\langle{\mathcal L_{ij} \mathcal M_{ij}}\rangle}{\langle{\mathcal M_{kl}\mathcal M_{kl}}\rangle},
\label{dsm_cs}
\end{equation} 
where $\mathcal L_{ij}=\widetilde{\bar{u}_i \bar{u}_j}-\tilde{\bar{u}}_i \tilde{\bar{u}}_j$, and $\mathcal M_{ij}=\tilde{\alpha}_{ij}-\beta_{ij}$. Here $\alpha_{ij}= 2 \Delta^2 |\bar{S}|\bar{S}_{ij}$, $\beta_{ij}= 2 \tilde{\Delta}^2 |\tilde{\bar{S}}|\tilde{\bar{S}}_{ij}$. An overvar denotes the grid filter at a scale $\Delta$, a tilde indicates a test filter coarser than the grid filter, and the variables with a tilde over the overbars denotes the quantities with double-filtering operation at scale $\tilde{\Delta}=2 \Delta$.

The dynamic mixed model (DMM) \cite{liu1994, shi2008} combines functional modeling and structure modeling for SGS stress, which includes a subgrid-viscosity part and a scale-similarity part. The model coefficients are determined dynamically by the Germano identity \cite{germano1991},
\begin{equation}
L_{ij}=T_{ij}-\tilde{\tau}_{ij},
\label{gi}
\end{equation} 
where a tilde indicates a test filter coarser than the grid filter, ${T_{ij}} = \widetilde {\overline {{u_i}{u_j}} } - {\tilde{\bar u}}_i{\tilde{\bar u}}_j$ is the SGS stress at the double-filtering scale $\tilde{\Delta}=2 \Delta$, and ${L_{ij}} = \widetilde {{{\bar u}_i}{{\bar u}_j}} - {\tilde{\bar u}}_i{\tilde{\bar u}}_j$ is the resolved stress. The deviatoric part of the SGS stress at scale $\Delta$ and $\tilde{\Delta}$ are modeled by \cite{shi2008,xie2018},
\begin{equation}
\tau_{ij}-\frac{\delta_{ij}}{3}\tau_{kk}={C_1}h_{1,ij}^A + {C_2}h_{2,ij}^A,
\label{dmm1}
\end{equation} 
\begin{equation}
T_{ij}-\frac{\delta_{ij}}{3}T_{kk}={C_1}H_{1,ij}^A + {C_2}H_{2,ij}^A,
\label{dmm2}
\end{equation}  
where $h_{1,ij}^A = -2{\Delta ^2}\left| {\bar S} \right|{\bar S_{ij}}$, $h_{2,ij}^A = {h_{2,ij}} - \frac{{{\delta _{ij}}}}{3}{h_{2,kk}}$, ${h_{2,ij}} = \widetilde {{{\bar u}_i}{{\bar u}_j}} - {\tilde {\bar u}_i}{\tilde {\bar u}_j}$, $H_{1,ij}^A = -2{\tilde \Delta ^2}\left| {\tilde {\bar S}} \right|{\tilde {\bar S}_{ij}}$, $H_{2,ij}^A = {H_{2,ij}} - \frac{{{\delta _{ij}}}}{3}{H_{2,kk}}$, and ${H_{2,ij}} = \widehat {{{\tilde {\bar u}}_i}{{\tilde {\bar u}}_j}} - {\hat {\tilde {\bar u}}_i}{\hat {\tilde {\bar u}}_j}$, the hat denotes the test filtering at scale $\hat{\Delta}=4\Delta$. The model coefficients $C_1$ and $C_2$ are determined by least squares algorithom \cite{shi2008},
\begin{equation}
{C_1} = \frac{{\left\langle {N_{ij}^2} \right\rangle \left\langle {L_{ij}^A{M_{ij}}} \right\rangle  - \left\langle {{M_{ij}}{N_{ij}}} \right\rangle \left\langle {L_{ij}^A{N_{ij}}} \right\rangle }}{{\left\langle {N_{ij}^2} \right\rangle \left\langle {M_{ij}^2} \right\rangle  - {{\left\langle {{M_{ij}}{N_{ij}}} \right\rangle }^2}}},
\label{dmm_c1}
\end{equation} 
 
\begin{equation}
{C_2} = \frac{{\left\langle {M_{ij}^2} \right\rangle \left\langle {L_{ij}^A{N_{ij}}} \right\rangle  - \left\langle {{M_{ij}}{N_{ij}}} \right\rangle \left\langle {L_{ij}^A{M_{ij}}} \right\rangle }}{{\left\langle {N_{ij}^2} \right\rangle \left\langle {M_{ij}^2} \right\rangle  - {{\left\langle {{M_{ij}}{N_{ij}}} \right\rangle }^2}}},
\label{dmm_c2}
\end{equation} 
where ${M_{ij}} = H_{1,ij}^A - \widetilde {h_{1,ij}^A}$, and ${N_{ij}} = H_{2,ij}^A - \widetilde {h_{2,ij}^A}$.

\begin{figure}\centering
	\begin{subfigure}{0.5\textwidth}
		\centering
		\includegraphics[width=0.9\linewidth]{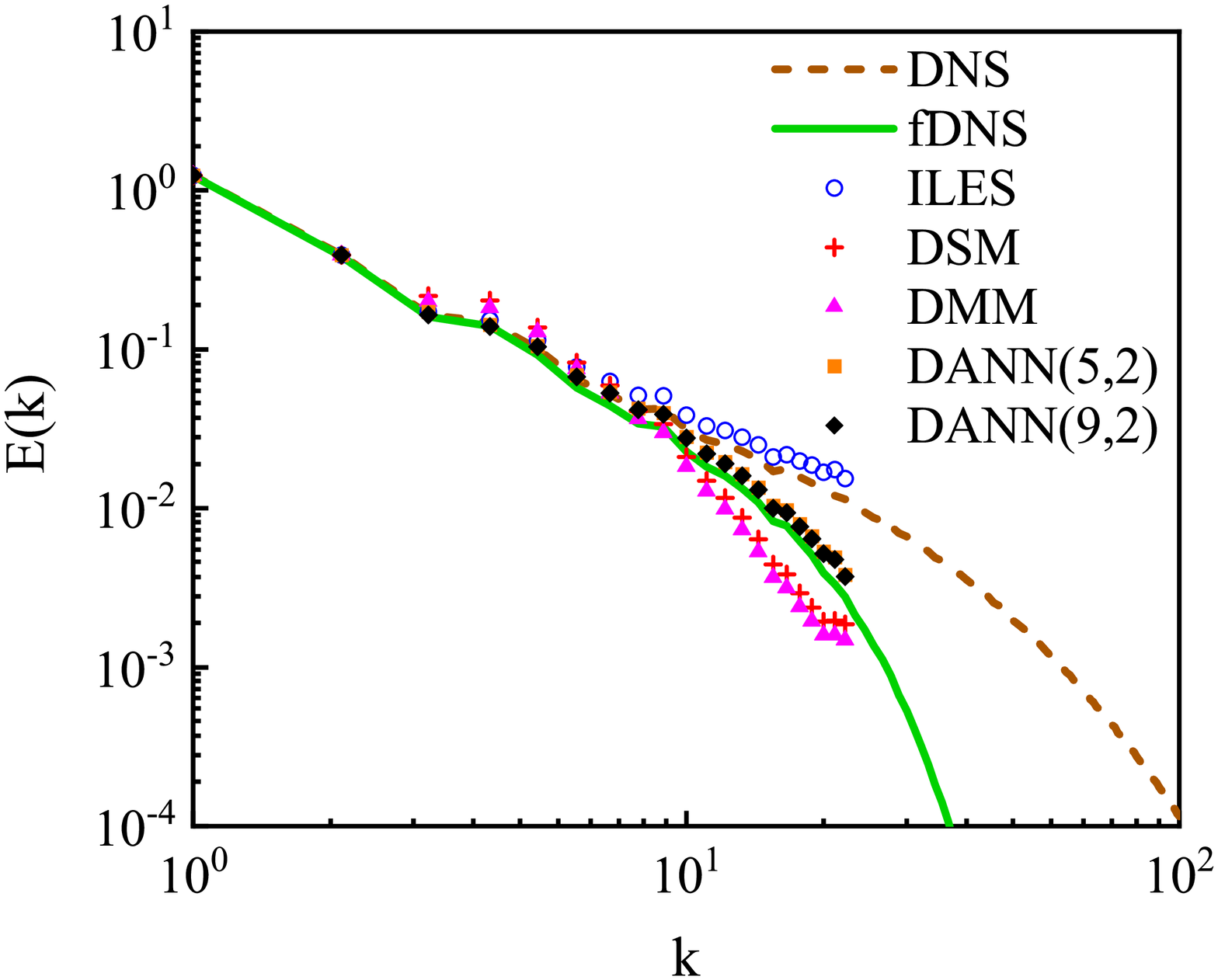}
		\caption{$\Delta_{LES}=\Delta/2(64^3)$}
	\end{subfigure}%
	\begin{subfigure}{0.5\textwidth}
		\centering
		\includegraphics[width=0.9\linewidth]{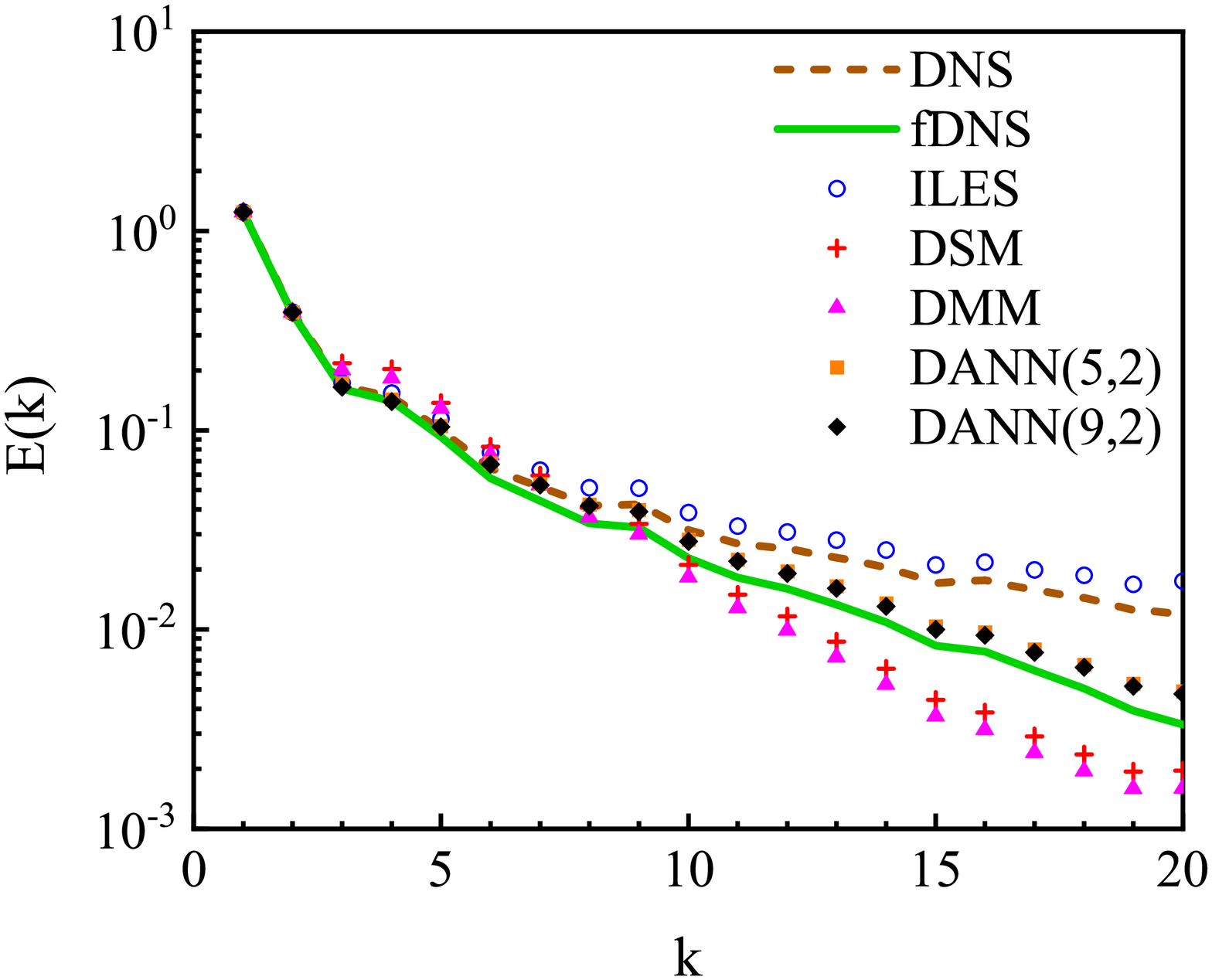}
		\caption{$\Delta_{LES}=\Delta/2(64^3)$ and k $\le $20}
	\end{subfigure}%
	
	\caption{ Comparisons of velocity spectrum for LES at grid resolution of $64^3$($\Delta_{LES}=\Delta/2$) with the same filter width $\Delta=32\Delta_{DNS}$.}\label{fig:6}
\end{figure}

The \emph{a posterior} performances of the DANN(5,2) model at the grid resolution of $64^3$ ($\Delta_{LES}=\Delta/2$) with the filter width $\Delta=32\Delta_{DNS}$ are evaluated by the energy spectrum and statistics of the velocity. The velocity spectra of LES with no-model(ILES), DSM, DMM, DANN(5,2) and DANN(9,2) models compared to those of DNS, filtered DNS(fDNS) data are shown in Fig.~\ref{fig:6}. The velocity spectrum of DNS data has a long inertial range that satisifies Kolmogorov hypothesis with a $k^{-5/3}$ scaling. The spectrum of fDNS data decays much faster than that of DNS data in the high wavenumber region, since the small-scale energy is filtered out. For different SGS models, the prediction errors become larger as the wavenumber k increases. The ILES model obviously overestimates the velocity spectrum at large wavenumbers. The energy spectra of the DSM and DMM models have the distinct tilted distribution, where energy at the low wavenumber region is enriched, while that near the cutoff wavenumber is excessively dissipated. Compared to these classical SGS models, the velocity spectrum predicted by the DANN(5,2) model is almost overlapped with that of the fDNS data.
 
\begin{figure}\centering
	\includegraphics[width=0.7\textwidth]{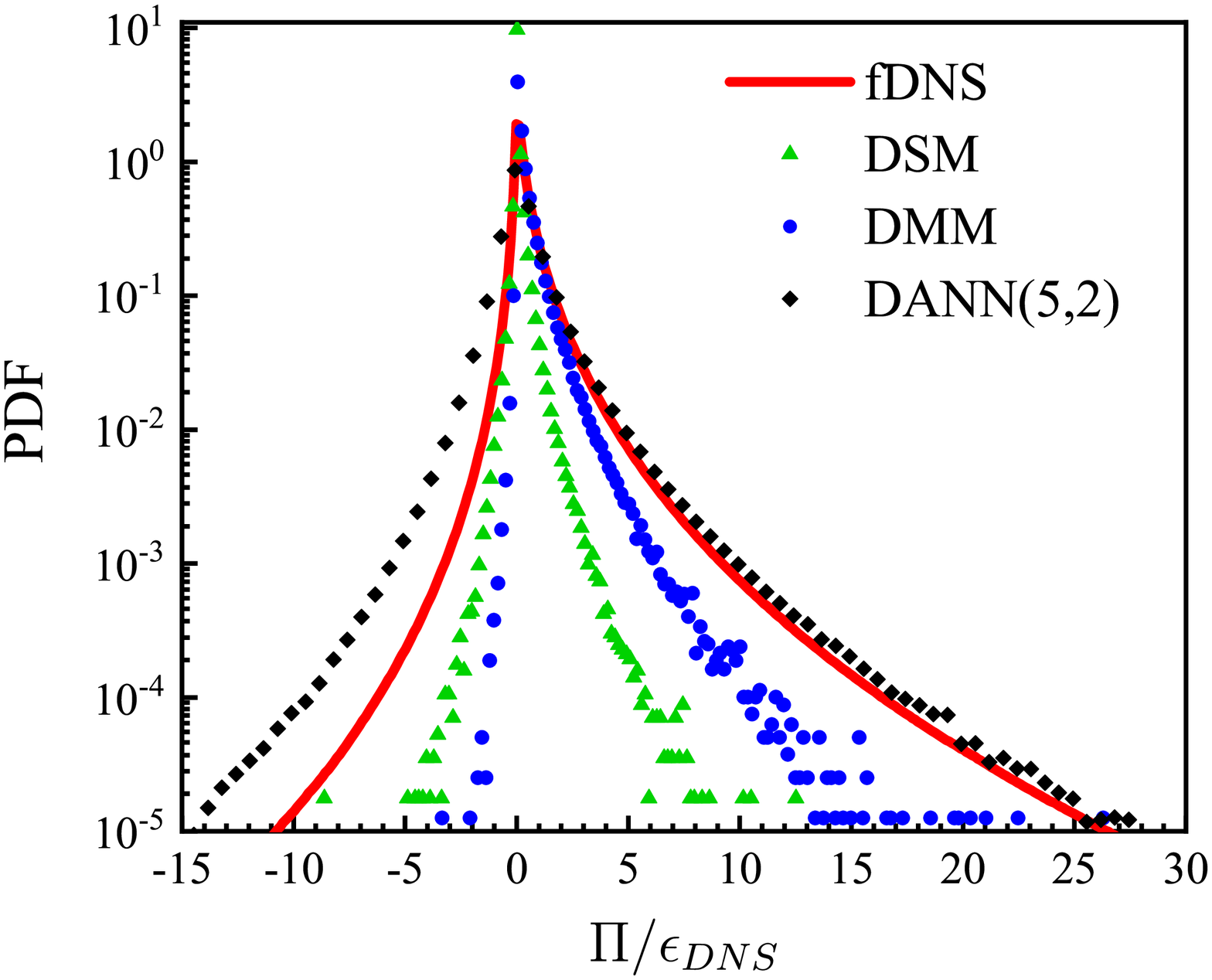}
	\caption{ PDFs of the normalized SGS flux $\Pi/\epsilon_{DNS}$ for LES at grid resolution of $64^3$($\Delta_{LES}=\Delta/2$) with the filter width $\Delta=32\Delta_{DNS}$.}\label{fig:7}
\end{figure}

The SGS energy flux represents the kinetic energy transfer between filtered scales and residual scales. In order to compare with the fDNS data, the SGS flux is normalized by the energy dissipation rate of DNS data. The PDFs of the normalized SGS energy flux $\Pi/\epsilon_{DNS}$ for fDNS data and LES with the DSM, DMM, and DANN(5,2) models are displayed in Fig.~\ref{fig:7}. The DSM model underestimates the PDF of SGS flux and its PDF tails are shorter than those of the fDNS data. The DMM model improves the prediction of the right tail of the PDF. Compared to the DSM and DMM models, the PDF tails of the SGS energy flux predicted by the DANN(5,2) model are very close to those of fDNS data.

In order to further validate the DANN model in the predictions of multi-scale properties of turbulence, we calculate the longitudinal structure functions of velocity, which is given by \cite{xie2018,xie2020}
\begin{equation}
S_n\left( r \right) = \left\langle {{{\left| {\frac{{{\delta _r}\bar u}}{{{{\bar u}^{rms}}}}} \right|}^n}} \right\rangle,
\label{sf}
\end{equation} 
where n denotes the order of structure function, ${\bar u^{rms}} = {\left\langle {{{\bar u}_i}{{\bar u}_i}} \right\rangle ^{1/2}}$ is the rms filtered velocity, and ${\delta _r}\bar u = \left[ {{\bf{\bar u}}\left( {{\bf{x}} + {\bf{r}}} \right) - {\bf{\bar u}}\left( {\bf{x}} \right)} \right] \cdot {\bf{\hat r}}$ represents the longitudinal increment of the velocity at the separation $\bf{r}$. Here ${\bf{\hat r}} = {\bf{r}}/\left| {\bf{r}} \right|$ denotes the unit distance vector.
 
\begin{figure}\centering
	\begin{subfigure}{0.333\textwidth}
		\centering
		\includegraphics[width=0.9\linewidth]{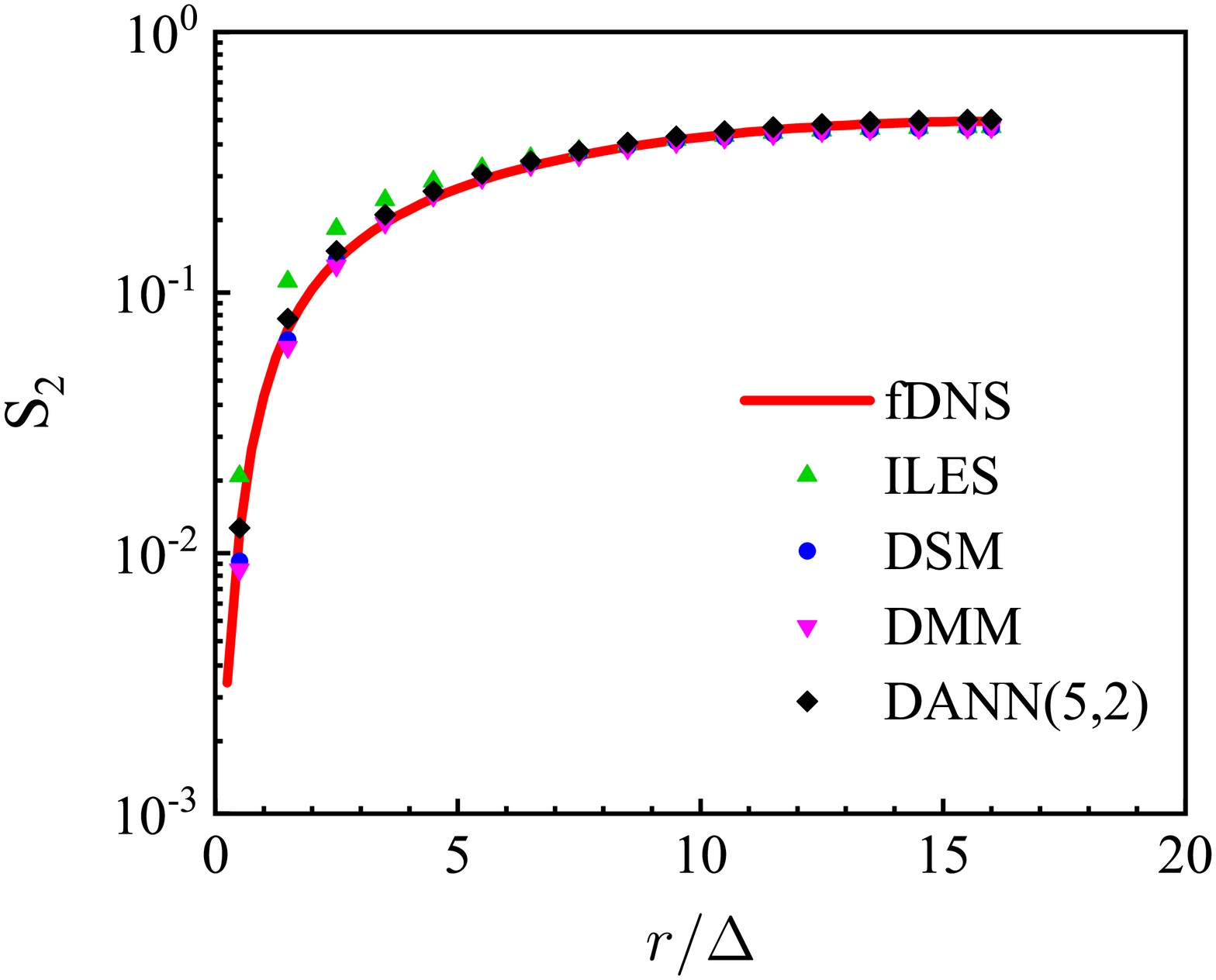}
		\caption{$S_2-r/\Delta$}
	\end{subfigure}%
	\begin{subfigure}{0.333\textwidth}
		\centering
		\includegraphics[width=0.9\linewidth]{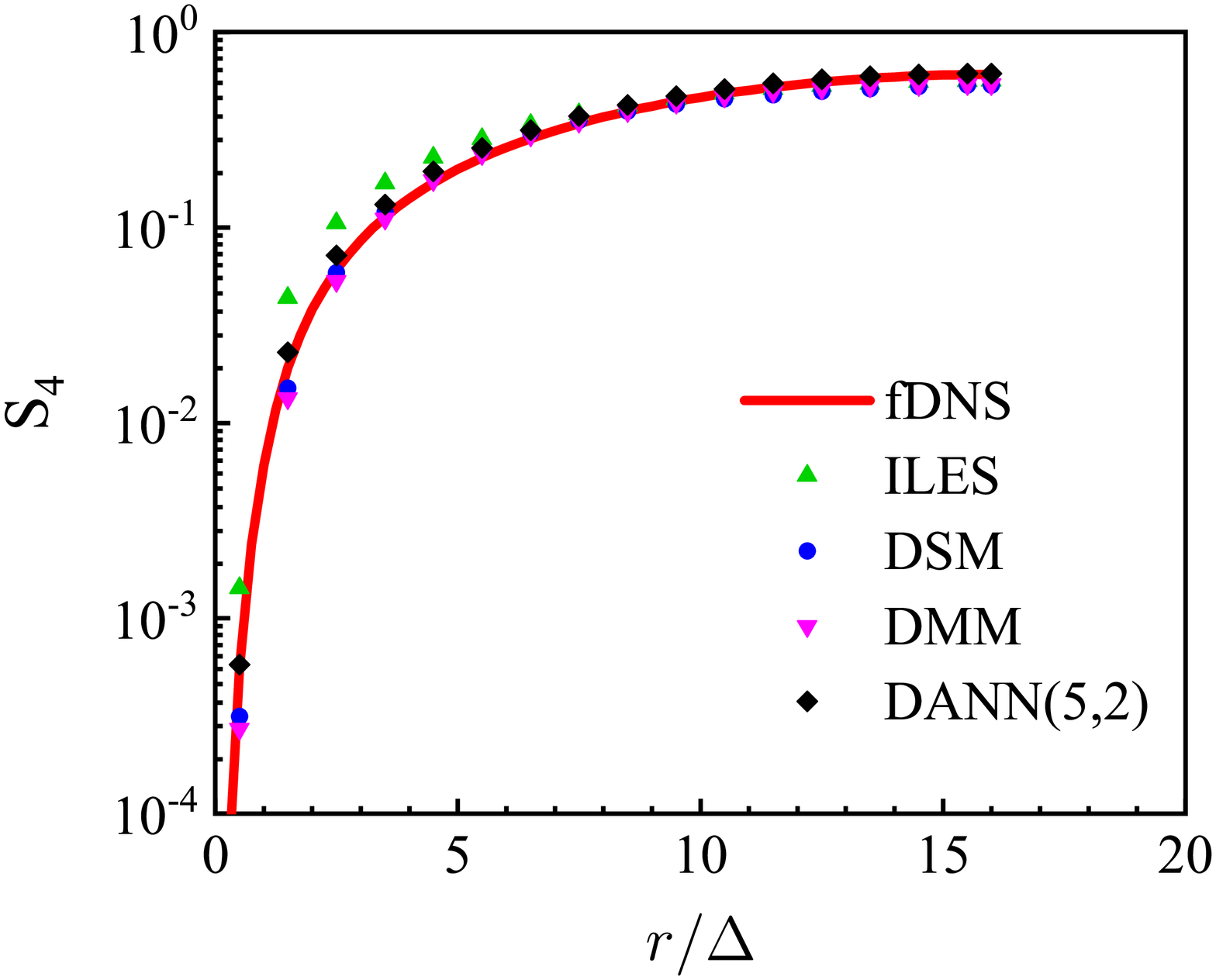}
		\caption{$S_4-r/\Delta$}
	\end{subfigure}%
	\begin{subfigure}{0.333\textwidth}
		\centering
		\includegraphics[width=0.9\linewidth]{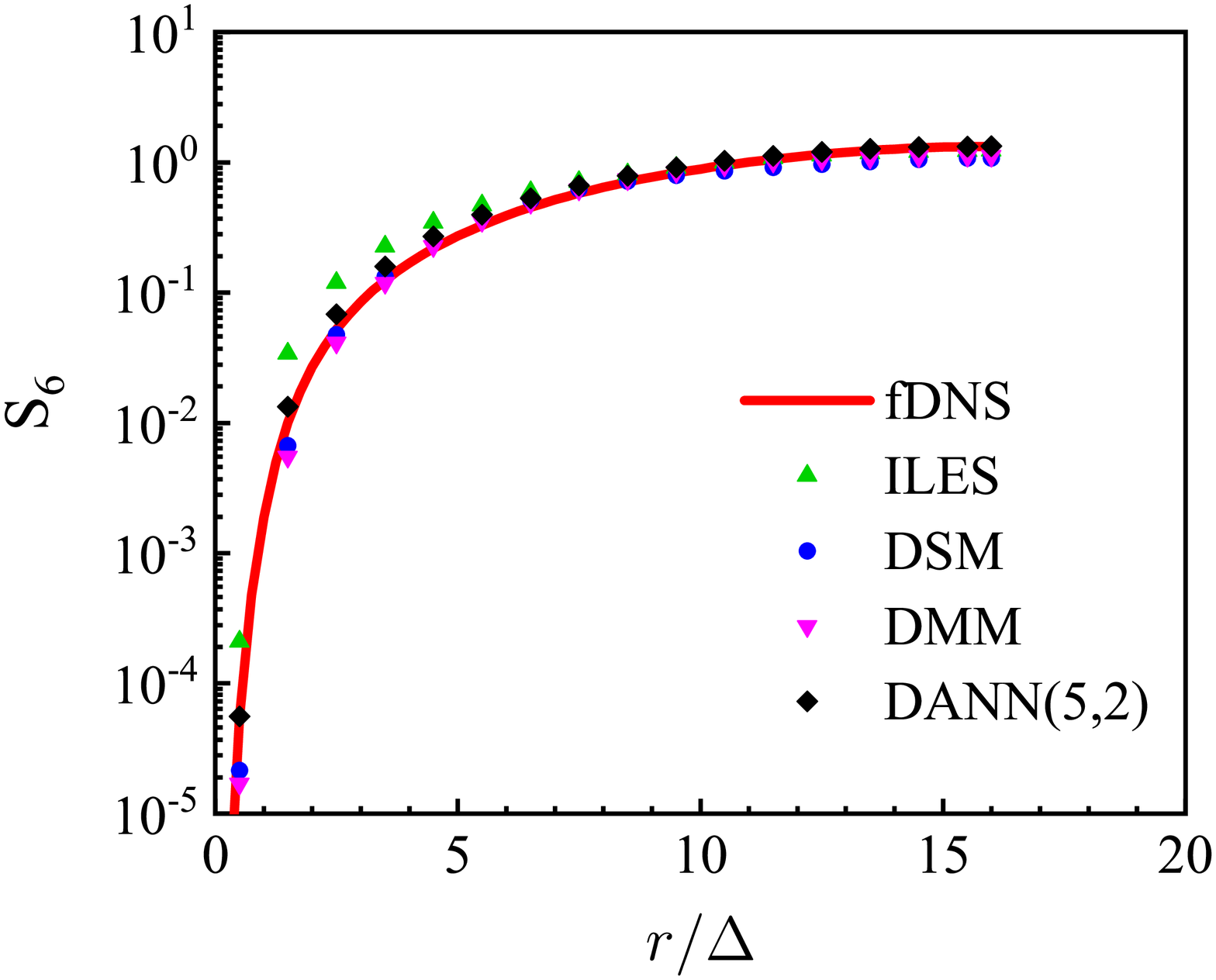}
		\caption{$S_6-r/\Delta$}
	\end{subfigure}%
	\caption{ Structure functions of the velocity for LES at grid resolution of $64^3$($\Delta_{LES}=\Delta/2$) with the same filter width $\Delta=32\Delta_{DNS}$ in the \emph{a posterior} study.}\label{fig:8}
\end{figure}

The comparisons of velocity structure functions predicted by LES with different SGS models are shown in Fig.~\ref{fig:8}. All SGS models can predict the velocity structure functions quite well at large separations $\bf{r}$. It can be clearly seen that the ILES model gives the worst prediction of the structure functions at small separations $\bf{r}$, which is significantly overestimated compared with those of the fDNS data. The predictions of the low-order structure functions ($S_2$ and $S_4$) by the DSM, DMM, and DANN(5,2) models are very close to each other. For the high-order structure function ($S_6$), the DANN(5,2) model predicts it accurately, while DMM and DSM models underestimate it at small separations. This result indicates that the DANN model can accurately predict velocity statistics of turbulence at different length scales.  
 
\begin{figure}\centering
	\begin{subfigure}{0.5\textwidth}
		\centering
		\includegraphics[width=0.9\linewidth]{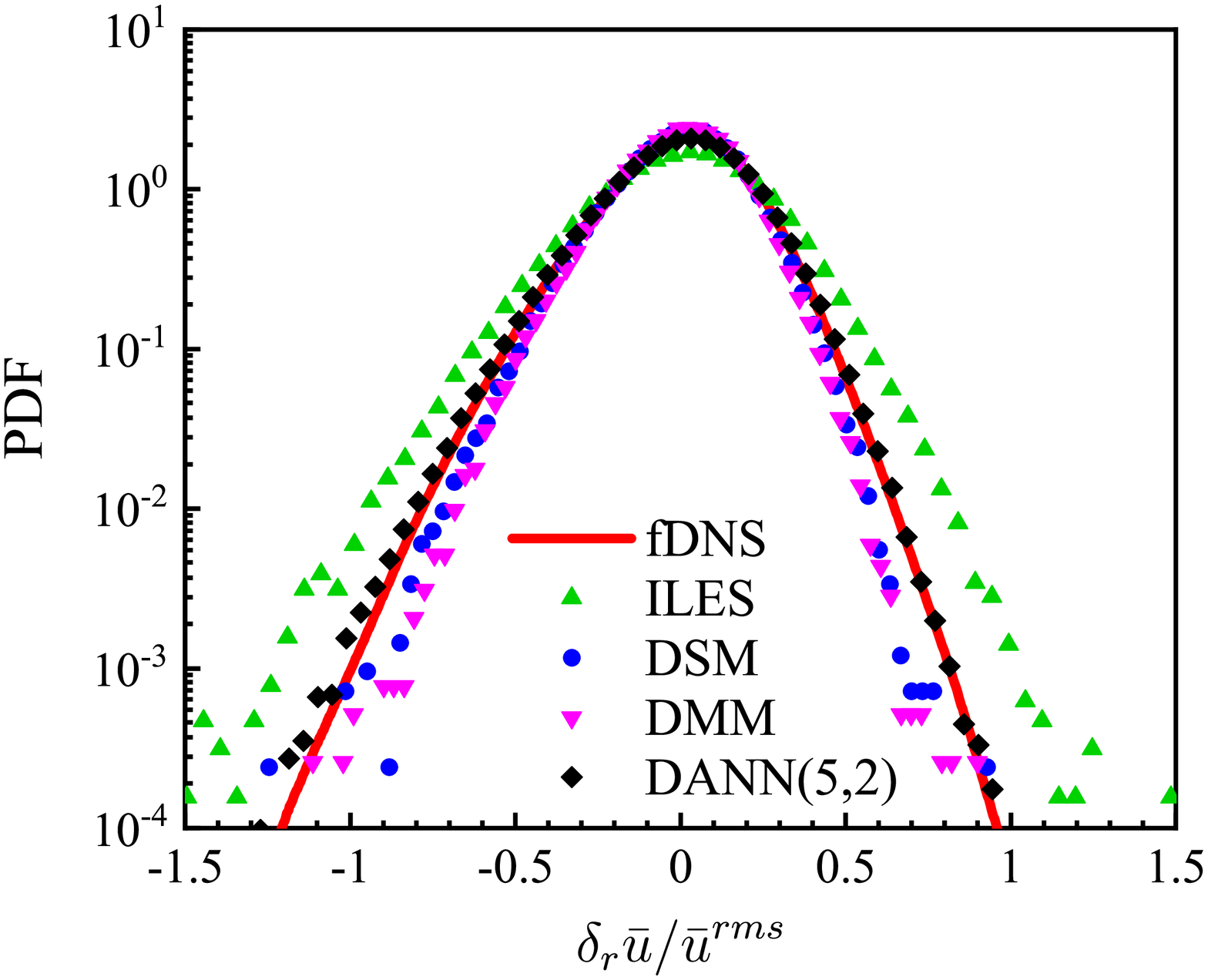}
		\caption{PDF of $\delta_r \bar u/\bar u^{rms}(r=\Delta)$}
	\end{subfigure}%
	\begin{subfigure}{0.5\textwidth}
		\centering
		\includegraphics[width=0.9\linewidth]{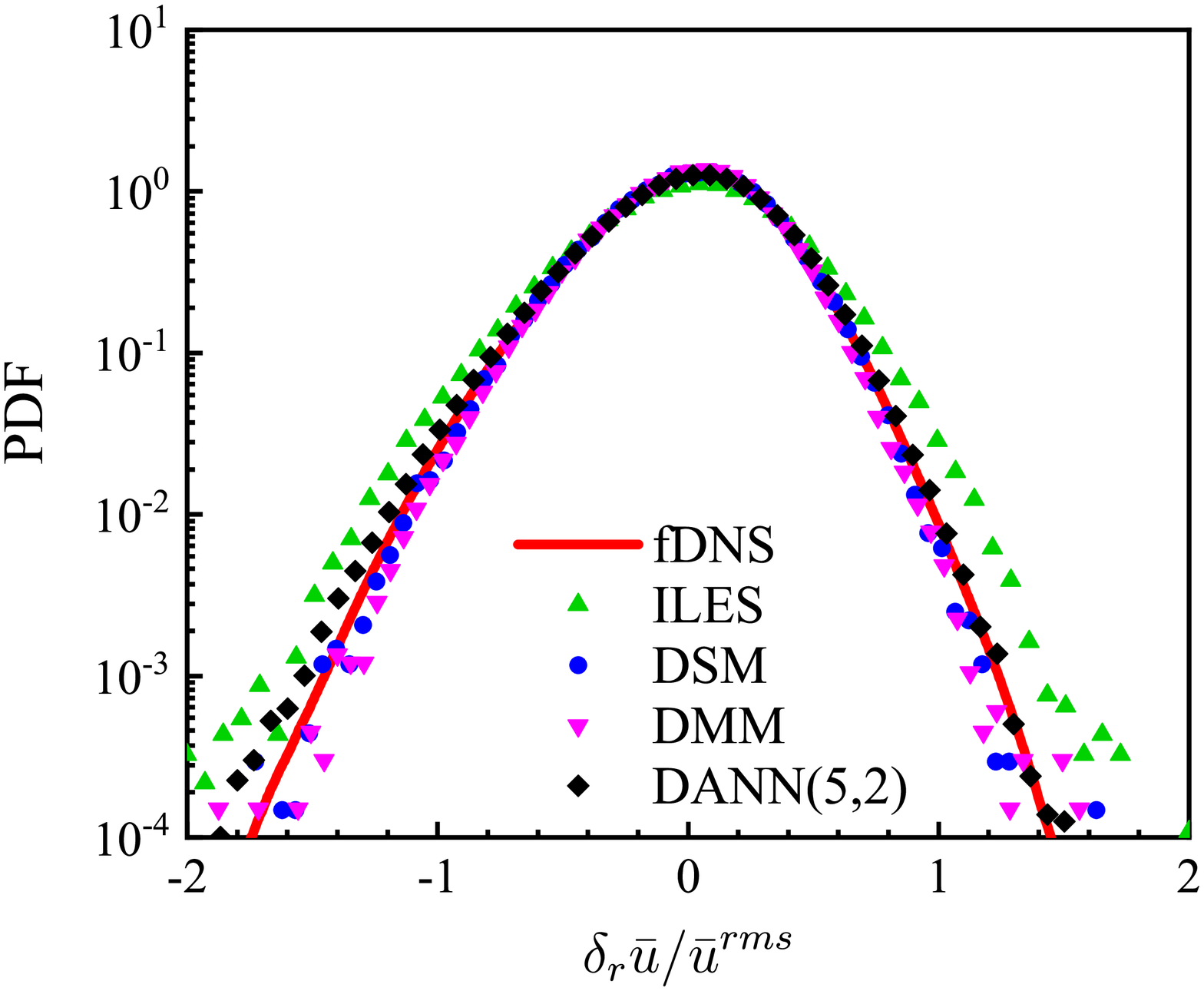}
		\caption{PDF of $\delta_r \bar u/\bar u^{rms}(r=2\Delta)$}
	\end{subfigure}\\
	\begin{subfigure}{0.5\textwidth}
		\centering
		\includegraphics[width=0.9\linewidth]{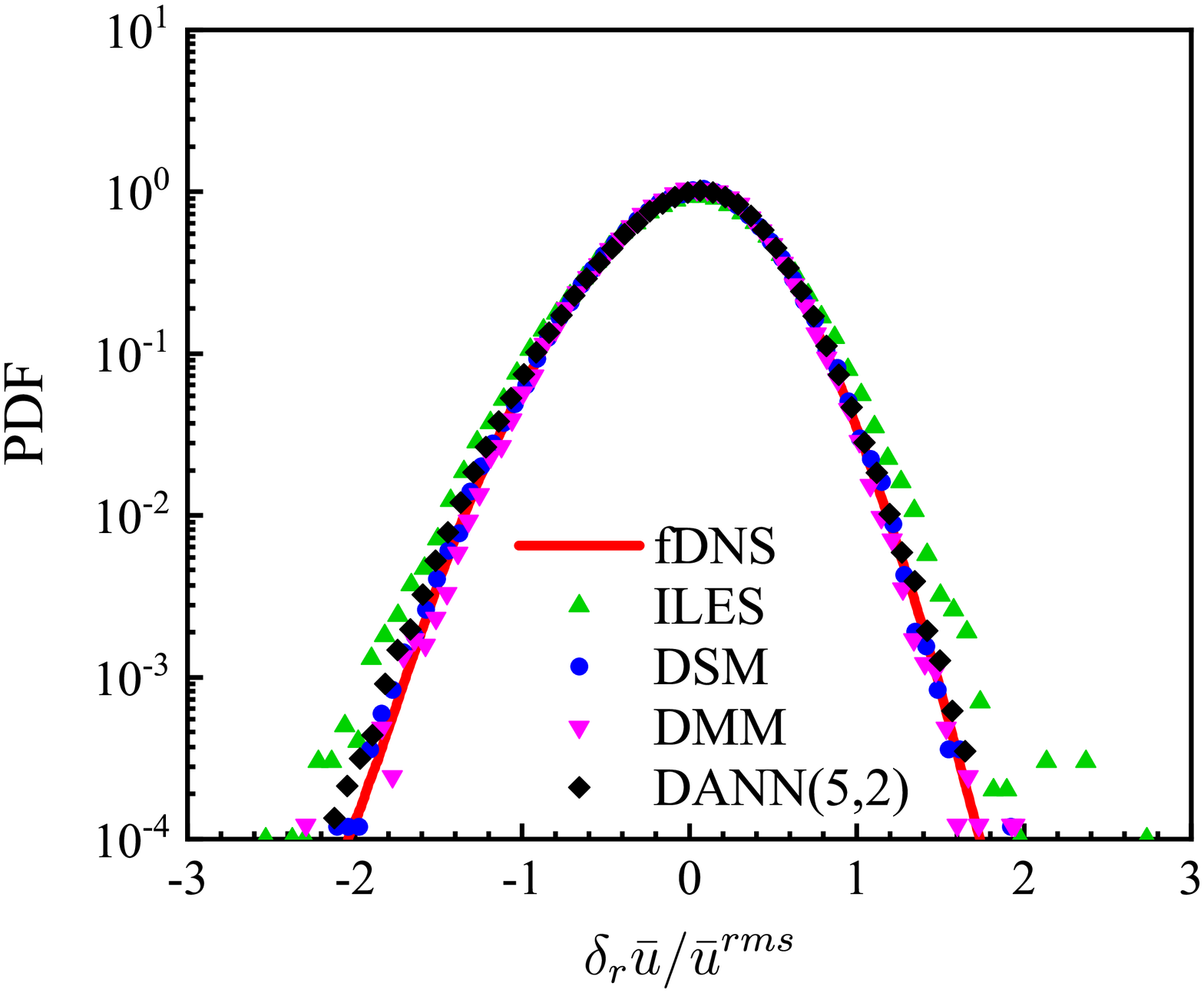}
		\caption{PDF of $\delta_r \bar u/\bar u^{rms}(r=3\Delta)$}
	\end{subfigure}%
	\begin{subfigure}{0.5\textwidth}
		\centering
		\includegraphics[width=0.9\linewidth]{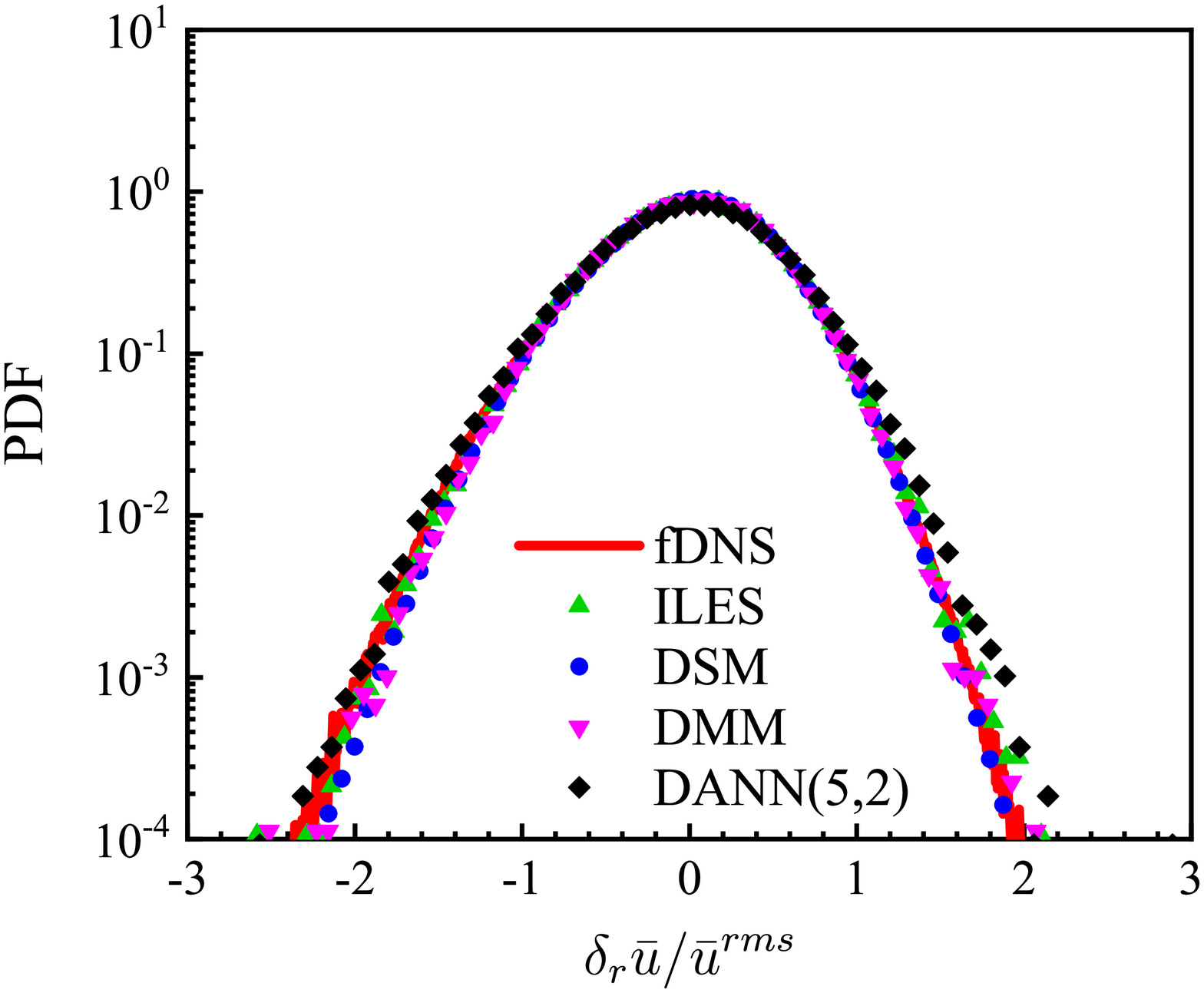}
		\caption{PDF of $\delta_r \bar u/\bar u^{rms}(r=4\Delta)$}
	\end{subfigure}%
	\caption{ PDFs of the normalized increments of the velocity for LES at grid resolution of $64^3$($\Delta_{LES}=\Delta/2$) with the filter width $\Delta=32\Delta_{DNS}$.}\label{fig:9}
\end{figure}

Furthermore, we compare PDFs of the normalized velocity increments with different distances $\bf{r}$ in Fig.~\ref{fig:9}. It is difficult for the SGS models to accurately predict the increments of velocity at smaller distances. At small distances, the PDFs of velocity increments predicted by the ILES are apparently wider, while those predicted by the DSM and DMM models are narrower, as compared to those of fDNS data. In contrast, the PDFs calculated by the DANN(5,2) model are in good agreement with those of fDNS. All LES models perform well for the predictions on the PDFs of increments of velocity at large distances. 

\begin{figure}\centering
	\begin{subfigure}{0.5\textwidth}
		\centering
		\includegraphics[width=0.9\linewidth]{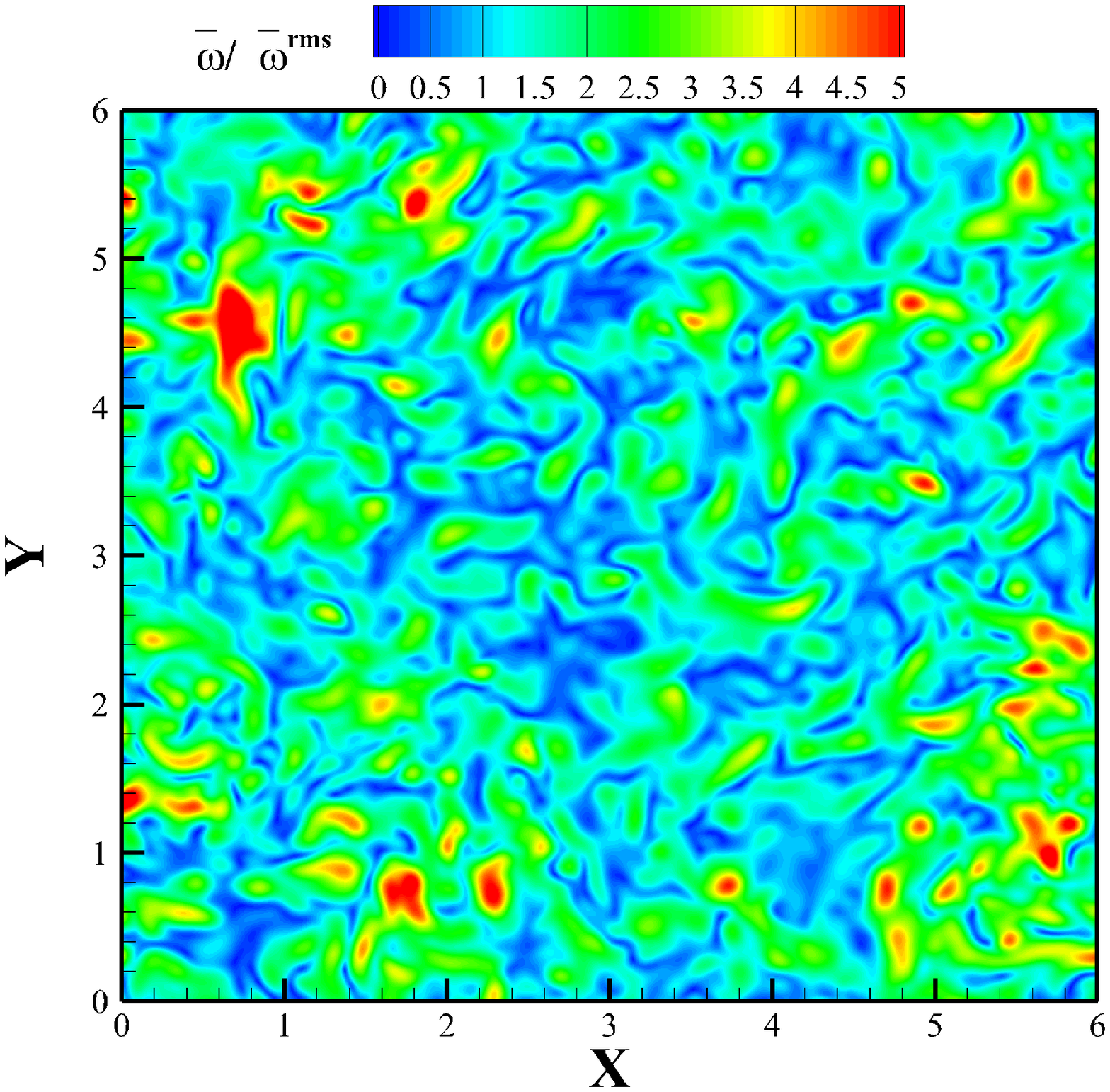}
		\caption{$\bar\omega/\bar\omega^{rms}$ for fDNS}
	\end{subfigure}%
	\begin{subfigure}{0.5\textwidth}
		\centering
		\includegraphics[width=0.9\linewidth]{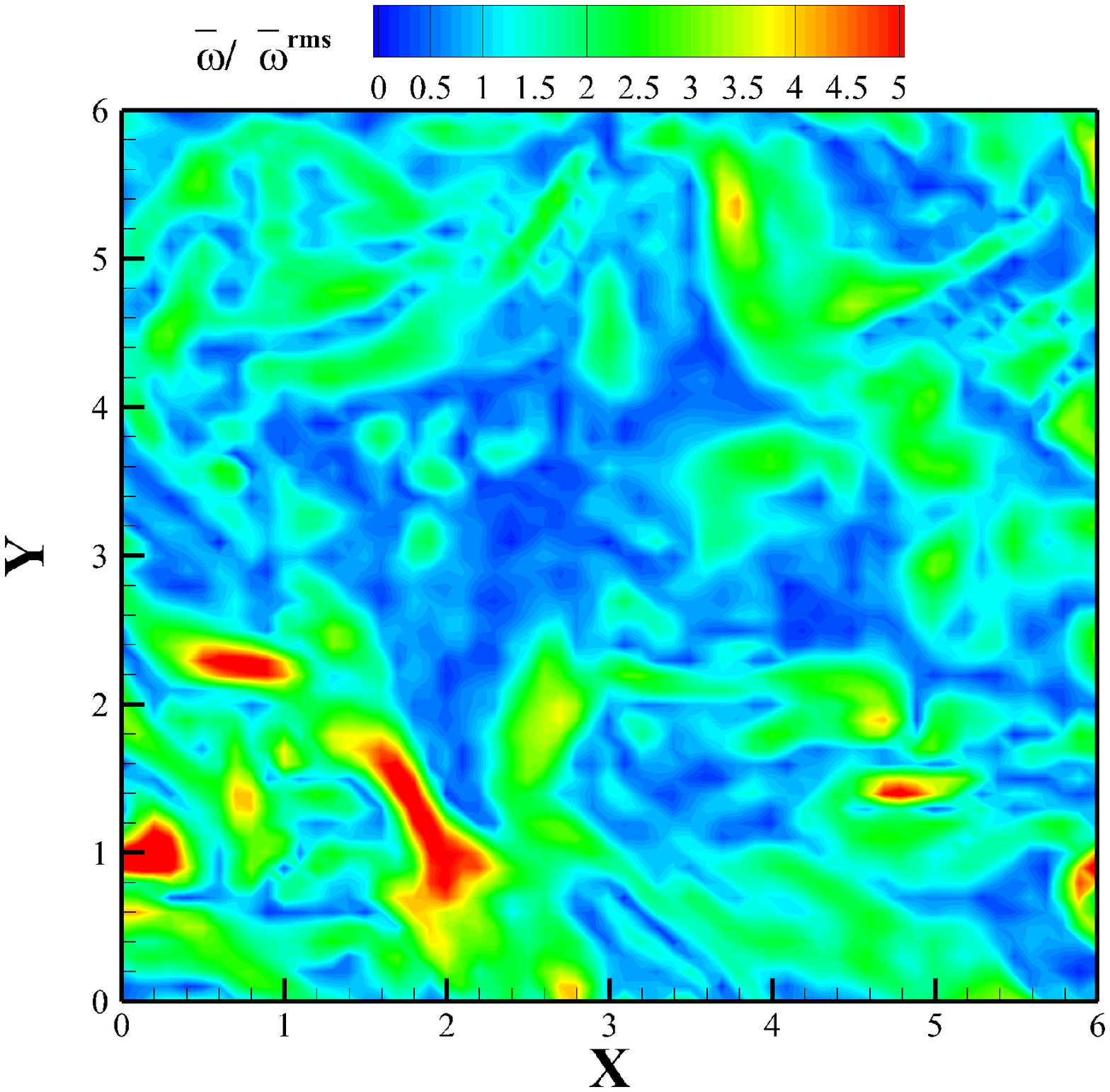}
		\caption{$\bar\omega/\bar\omega^{rms}$ for DSM}
	\end{subfigure}\\
	\begin{subfigure}{0.5\textwidth}
		\centering
		\includegraphics[width=0.9\linewidth]{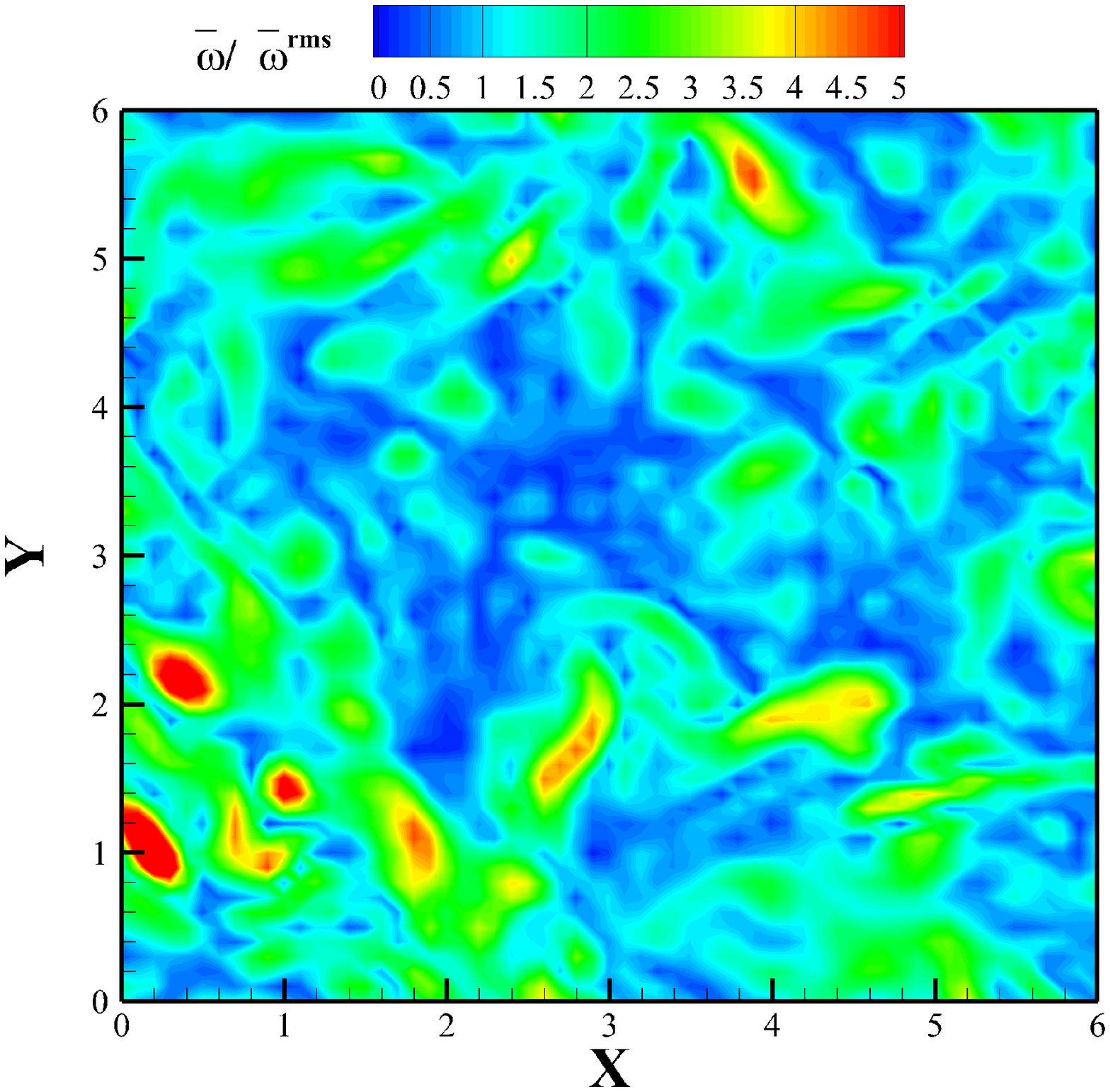}
		\caption{$\bar\omega/\bar\omega^{rms}$ for DMM}
	\end{subfigure}%
	\begin{subfigure}{0.5\textwidth}
		\centering
		\includegraphics[width=0.9\linewidth]{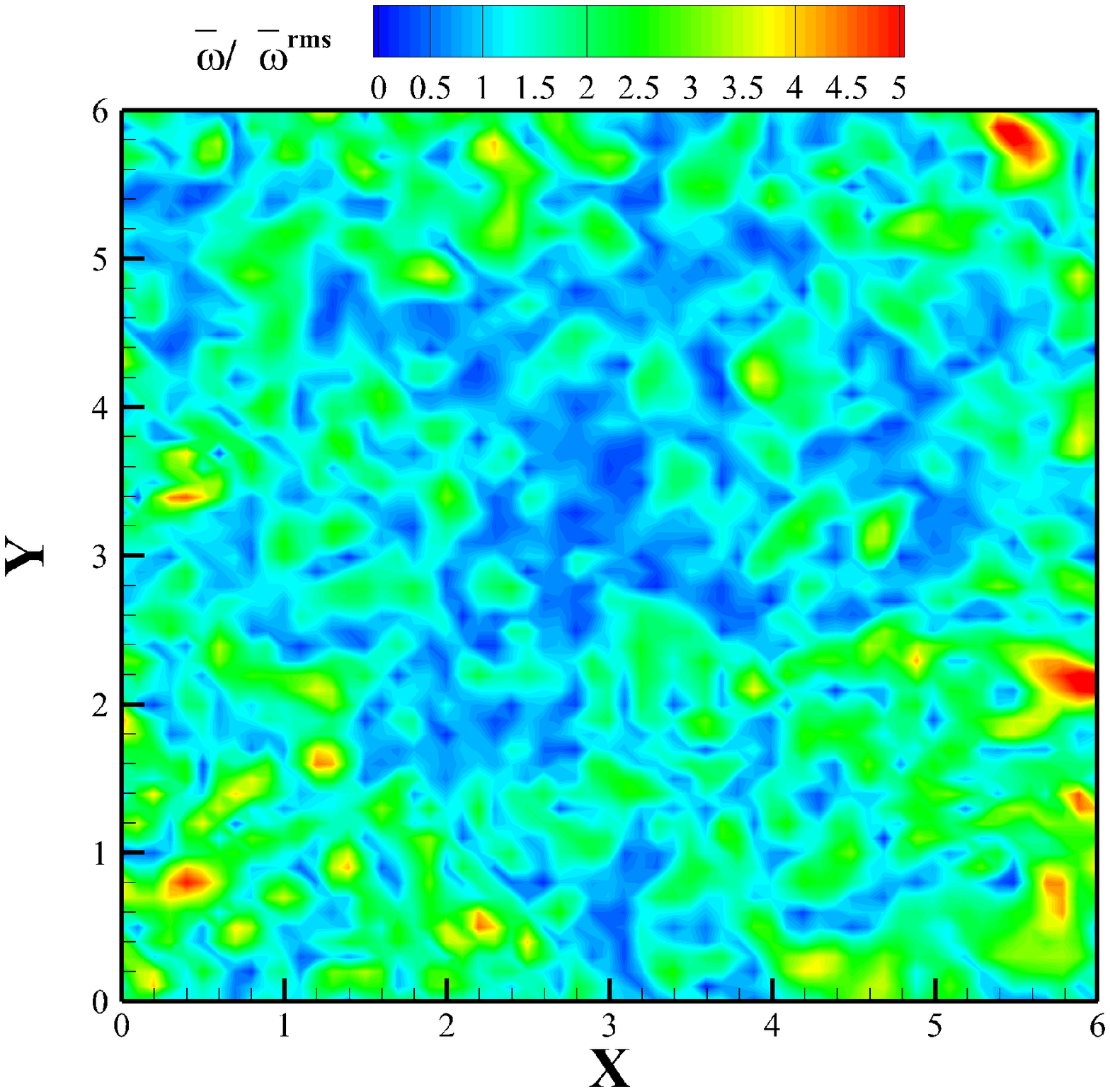}
		\caption{$\bar\omega/\bar\omega^{rms}$ for DANN(5,2)}
	\end{subfigure}%
	\caption{ Contours of the normalized vorticity $\bar\omega/\bar\omega^{rms}$ at arbitrary   x-y plane  at $t/\tau$=5 for LES at grid resolution of $64^3$ ($\Delta_{LES}=\Delta/2$) with the filter width $\Delta=32\Delta_{DNS}$.}\label{fig:10}
\end{figure}

The spatial distribution of the turbulent coherent structure can be examined by the instantaneous normalized vorticity contours \cite{xie2020a}. The LES computations with different SGS models are consistently initialized by the instantaneous flow field of the same fDNS data. Comparisons of the normalized vorticity contours at dimensionless time $t/\tau$=5 are displayed in Fig.~\ref{fig:10}, where $\tau=L_I/u^{rms}$ is the large eddy turnover time. All LES models (DSM, DMM and DANN(5,2) models) predict the vorticity contours quite well. It can be seen that some small-scale vortex structures are missing for the predictions of the DSM and DMM models, due to the excessive dissipation of these models. In comparison, the instantaneous vorticity contour predicted by the DANN(5,2) model catches more small-scale fluctuations and is closer to the fDNS data. 

\subsection{Validation at different filter widths ($\Delta=16\Delta_{DNS}$ and $64\Delta_{DNS}$)}
In order to test the generalization ability of the DANN models , we further analyze the performance of the trained DANN models at the filter widths different from the training filter width. The DANN models are trained by the fDNS data with filter width $\Delta=32\Delta_{DNS}$, which are validated at grid resolutions of $128^3$ and $32^3$ ($\Delta_{LES}=\Delta/2$) with filter width $\Delta=16\Delta_{DNS}$ and $64\Delta_{DNS}$, respectively. The DANN(5,2) model is selected to investigate the performance of the DANN models in \emph{a posteriori} study. The time step for LES at grid resolutions of $128^3$ and $32^3$ is $\Delta t_{LES}=8\Delta t_{DNS}$, which is consistent with that of grid resolution of $64^3$.

\begin{figure}\centering
	\begin{subfigure}{0.5\textwidth}
		\centering
		\includegraphics[width=0.9\linewidth]{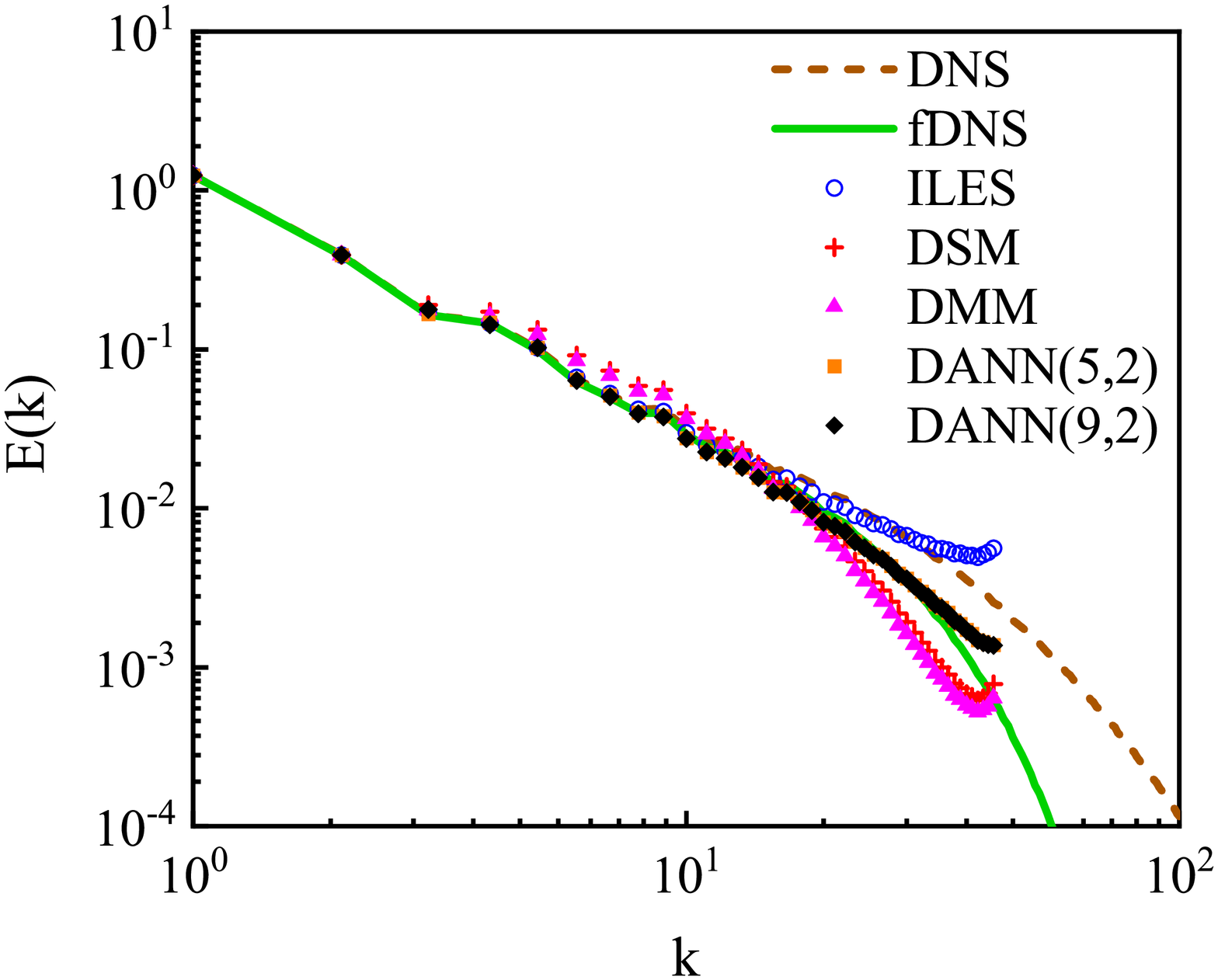}
		\caption{$\Delta_{LES}=\Delta/2(128^3)$}
	\end{subfigure}%
	\begin{subfigure}{0.5\textwidth}
		\centering
		\includegraphics[width=0.9\linewidth]{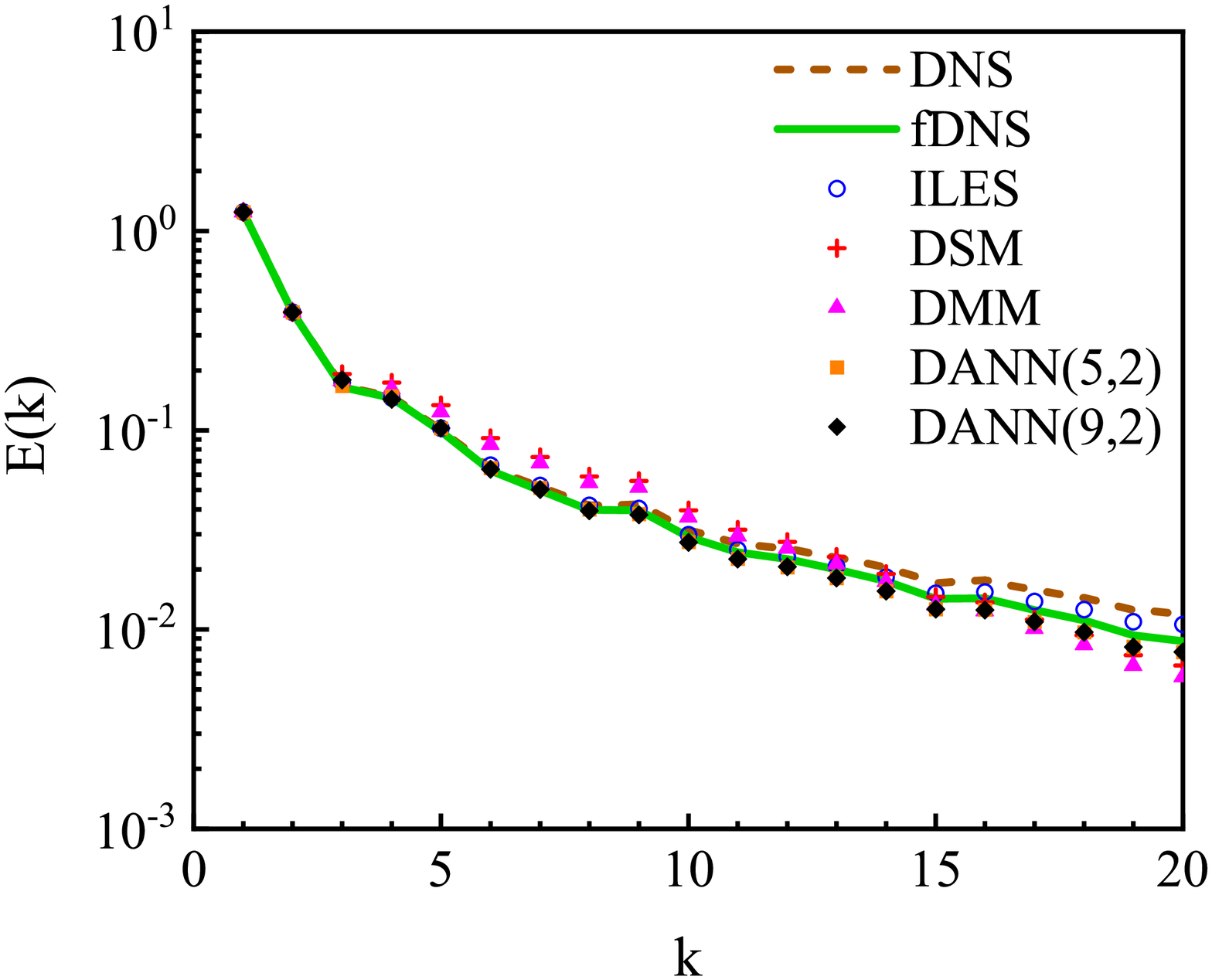}
		\caption{$\Delta_{LES}=\Delta/2(128^3)$ and k $\le $20}
	\end{subfigure}%
	
	\caption{Comparisons of velocity spectrum for LES at grid resolution of $128^3$($\Delta_{LES}=\Delta/2$) with the filter width $\Delta=16\Delta_{DNS}$.}\label{fig:11}
\end{figure}

\begin{figure}\centering
	\begin{subfigure}{0.5\textwidth}
		\centering
		\includegraphics[width=0.9\linewidth]{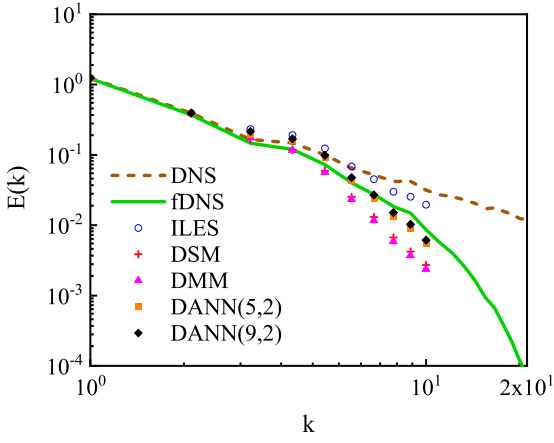}
		\caption{$\Delta_{LES}=\Delta/2(32^3)$}
	\end{subfigure}%
	\begin{subfigure}{0.5\textwidth}
		\centering
		\includegraphics[width=0.9\linewidth]{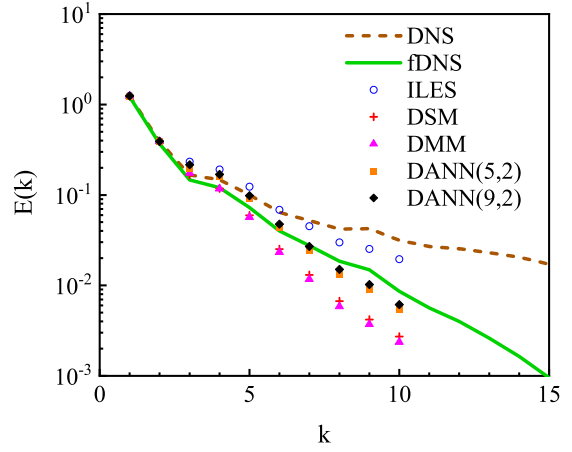}
		\caption{$\Delta_{LES}=\Delta/2(32^3)$ and k $\le $15}
	\end{subfigure}%
	
	\caption{Comparisons of velocity spectrum for LES at grid resolution of $32^3$($\Delta_{LES}=\Delta/2$) with the filter width $\Delta=64\Delta_{DNS}$. }\label{fig:12}
\end{figure}

The comparisons of velocity spectra with filter widths $\Delta=16\Delta_{DNS}$ and $\Delta=64\Delta_{DNS}$ for different SGS models (ILES, DSM, DMM and DANN(5,2)) are shown in Fig.~\ref{fig:11} and Fig.~\ref{fig:12}, respectively. The ILES model gives the worst prediction of the velocity spectrum, while the DSM and DMM models are too dissipative at high wavenumber. In comparison, the spectrum predicted by the DANN(5,2) model is in reasonable agreement with the fDNS data.

\begin{figure}\centering
	\begin{subfigure}{0.5\textwidth}
		\centering
		\includegraphics[width=0.9\linewidth]{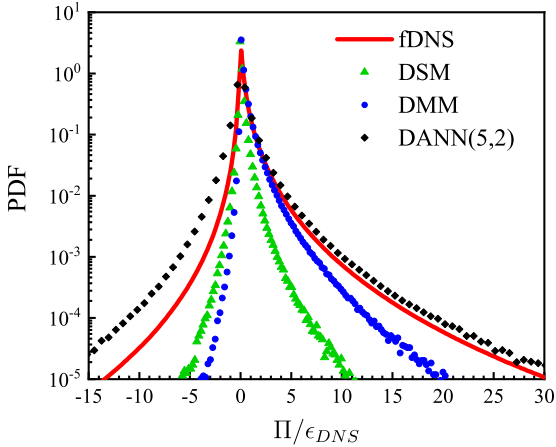}
		\caption{$\Pi/\epsilon_{DNS}$ at $128^3$ with $\Delta=16\Delta_{DNS}$}
	\end{subfigure}%
	\begin{subfigure}{0.5\textwidth}
		\centering
		\includegraphics[width=0.9\linewidth]{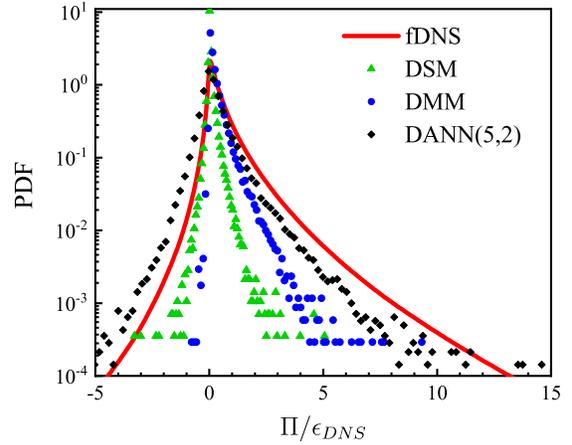}
		\caption{$\Pi/\epsilon_{DNS}$ at $32^3$ with $\Delta=64\Delta_{DNS}$}
	\end{subfigure}%
	\caption{PDFs of the normalized SGS flux $\Pi/\epsilon_{DNS}$ for LES at grid resolutions of $128^3$ and $32^3$($\Delta_{LES}=\Delta/2$) with the filter widths $\Delta=16\Delta_{DNS}$ and $\Delta=64\Delta_{DNS}$.}\label{fig:13}
\end{figure}

The PDFs of the normalized SGS flux $\Pi/\epsilon_{DNS}$ at grid resolutions of $128^3$ and $32^3$ with $\Delta=16\Delta_{DNS}$ and $64\Delta_{DNS}$ are shown in Fig.~\ref{fig:13}. The values of SGS energy flux magnitude are statistically underestimated by the DSM and DMM models, and the corresponding PDFs are significantly narrower than that of the fDNS data. In contrast to these traditional SGS models, the DANN(5,2) model gives the better shape of PDF of SGS energy flux which is quite close to that of the fDNS data.

\section{Conclusions} \label{sec:level6}
To summarize, a deconvolutional artificial neural network (DANN) framework is proposed to model the SGS stress in LES of turbulence. The accuracy of the DANN models can be improved by increasing the spacing width D and the ratio of stencil size $\Delta/\Delta_L$, due to the incorporation of more spatial features of the flow field at the neighboring points. Compared to the classical ADM and VGM models, the DANN models can recover the SGS stress more accurately with the correlation coefficient of up to 99\% and the relative error of less than 15\% in the \emph{a priori} analysis. Compared to the ADM models, the spectra of the unfiltered velocity predicted by the DANN models are closer to the results of the DNS data. The PDFs of normalized SGS energy flux reconstructed by the DANN models are also more accurate than the traditional VGM and ADM models in the \emph{a prior} study. We further systematically evaluate the performance of the DANN(5,2) model by comparing the velocity spectra, PDF of SGS energy flux, the statistical properties of the velocity, as well as the instantaneous spatial structures of vorticity with the conventional SGS models (ILES, DSM, and DMM) in the \emph{a posteriori} study. The results indicate that the ILES model fails to accurately predict the statistical properties of turbulence, while the DSM and DMM models are too dissipative to recover the small-scale structures. In contrast, the DANN(5,2) model can not only predict the spectrum and statistics of velocity accurately, but also reconstruct the instantaneous coherent structures of vorticity correctly without increasing the considerable computational cost. Furthermore, we also validate the general performance of the DANN models by changing the filter width in the \emph{a posteriori} tests. The predictions given by trained DANN models without fine-tuning are better than the classical SGS models. 

However, several issues need further study. These include the incorporations of more \emph{a priori} knowledge and physical constraints into SGS models, the interpretability and universality of artificial neural networks, and the spatial-temporal correlations of SGS fields.

\acknowledgments{This work was supported by the National Numerical Windtunnel Project, by the National Natural Science Foundation of China (NSFC Grants No. 91952104, No. 11702127 and No. 91752201), by the Technology and Innovation Commission of Shenzhen Municipality (Grant Nos. KQTD20180411143441009 and JCYJ20170412151759222), and by Department of Science and Technology of Guangdong Province (Grant No. 2019B21203001). This work was also funded by the Center for Computational Science and Engineering of the Southern University of Science and Technology. J. W. particularly thanks the support from the Young Elite Scientist Sponsorship Program by CAST (Grant No. 2016QNRC001).}

\section{DATA AVAILABILITY} \label{sec:level7}
The data that support the findings of this study are available from the corresponding
author upon reasonable request.

\bibliography{reference}
\end{document}